\documentclass[twocolumn,english,aps,prl,showpacs,superscriptaddress,nobibnotes,longbibliography]{revtex4-1}
\usepackage[T1]{fontenc}
\usepackage[latin9]{inputenc}
\usepackage{units}
\usepackage{amsmath}
\usepackage{amssymb}
\usepackage{graphicx}

\makeatletter
\usepackage[breaklinks]{hyperref}
\usepackage{breakurl}

\makeatother

\usepackage{babel}
\begin{document}
\title{Rotation sensing with improved stability using point source atom interferometry}
\author{Chen Avinadav}
\thanks{These authors contributed equally to this work.~\\
chen.avinadav@weizmann.ac.il~\\
dimitry.yankelev@weizmann.ac.il}
\affiliation{Department of Physics of Complex Systems, Weizmann Institute of Science,
Rehovot 7610001, Israel}
\affiliation{Rafael Ltd, Haifa 3102102, Israel}
\author{Dimitry Yankelev}
\thanks{These authors contributed equally to this work.~\\
chen.avinadav@weizmann.ac.il~\\
dimitry.yankelev@weizmann.ac.il}
\affiliation{Department of Physics of Complex Systems, Weizmann Institute of Science,
Rehovot 7610001, Israel}
\affiliation{Rafael Ltd, Haifa 3102102, Israel}
\author{Moshe Shuker}
\affiliation{Rafael Ltd, Haifa 3102102, Israel}
\author{Ofer Firstenberg}
\affiliation{Department of Physics of Complex Systems, Weizmann Institute of Science,
Rehovot 7610001, Israel}
\author{Nir Davidson}
\affiliation{Department of Physics of Complex Systems, Weizmann Institute of Science,
Rehovot 7610001, Israel}
\begin{abstract}
Point source atom interferometry is a promising approach for implementing
robust, high-sensitivity, rotation sensors using cold atoms. However,
its scale factor, \emph{i.e.}, the ratio between the interferometer
signal and the actual rotation rate, depends on the initial conditions
of the atomic cloud, which may drift in time and result in bias instability,
particularly in compact devices with short interrogation times. We
present two methods to stabilize the scale factor, one relying on
a model-based correction which exploits correlations between multiple
features of the interferometer output and works on a single-shot basis,
and the other a self-calibrating method where a known bias rotation
is applied to every other measurement, requiring no prior knowledge
of the underlying model but reducing the sensor bandwidth by a factor
of two. We demonstrate both schemes experimentally with complete suppression
of scale factor drifts, maintaining the original rotation sensitivity
and allowing for bias-free operation over several hours.
\end{abstract}
\maketitle

\section{Introduction}

Cold atom interferometers \citep{Tino2014} have achieved in recent
years record sensitivities in acceleration and rotation sensing. As
acceleration-sensing instruments, their applications range from precision
measurements for fundamental research \citep{Weiss1993,Dimopoulos2007,Mueller2010,Bouchendira2011,Rosi2014,Zhou2015a,Barrett2016,Parker2018,Becker2018}
to geophysical measurements with mobile atomic gravimeters \citep{Bongs2019,Wu2009,Farah2014,Freier2016,Menoret2018,Bidel2018,Wu2019,Bidel2019},
demonstrating both high sensitivity and high stability operation.
Atom interferometry gyroscopes \citep{Barrett2014,Gustavson1997,Stockton2011,Dickerson2013,Savoie2018,Chen2019}
are useful for field applications such as gyrocompassing \citep{Gauguet2009,Sugarbaker2013}
and inertial navigation on mobile platforms \citep{JEKELI2005,Canuel2006,Rakholia2014}.
Similarly to atomic and optical clocks, atomic gyroscopes hold the
promise of enhanced stability compared to their classical counterparts,
due to their scale factor being defined in terms of fundamental physical
constants. Previous demonstrations of atom interferometry gyroscopes
\citep{Durfee2006,Savoie2018} have reached sensitivity and stability
which compare favorably with state-of-the-art optical gyroscopes \citep{Lefevre2014,Battelier2016}.

Point source interferometry (PSI) is an atom interferometry technique
for rotation sensing based on detecting the spatial frequency of the
interference pattern across an atomic cloud. Originally developed
in a 10-meter atomic fountain \citep{Dickerson2013}, the technique
has also been applied in a sensor with cm-scale physics package \citep{Hoth2016}.
Compared to other atom interferometry rotation sensing techniques,
PSI has the benefits of experimental simplicity and inherent suppression
of accelerations and vibrations. As such, it is an especially promising
technique for field and mobile applications. However, unlike ideal
atomic sensors, the scale factor of PSI, defined as the ratio between
the measured spatial frequency of the fringe pattern and the applied
rotation rate, is sensitive to the initial and final size of the atomic
cloud \citep{Hoth2017}. This dependency is amplified when the expansion
ratio of the cloud is small, as in compact sensors with short expansion
times. As a result, PSI is susceptible to bias instability due to
drifts in its technical attributes \citep{Chen2019}, preventing it
from realizing its full potential as an atomic sensor and limiting
its usefulness in applications requiring high stability over long
time scales.

In this work, we introduce two approaches to stabilizing scale-factor
drifts in PSI sensors. The first approach utilizes additional information
extracted from each PSI image, namely the interference fringe contrast
and atomic cloud final size, in addition to the fringe spatial frequency.
We experimentally verify the physical model which describes the correlation
between these parameters and then employ it to correct the scale factor
independently for every PSI image. The second approach relies on alternately
applying a known bias rotation rate to the sensor, in addition to
the unknown measured rotation. Analyzing pairs of measurements with
and without the bias rotation enables self-calibration of the scale
factor correction and determination of the unknown rotation. We implement
both schemes experimentally and demonstrate their ability to recover
uncorrelated $\tau^{-1/2}$ averaging performance of the rotation
sensor. We achieve suppression of scale factor drifts up to a factor
of ten without any loss in sensitivity and on time scales of up to
$\unit[10^{4}]{sec}$, far surpassing previous results.

\begin{figure}[t]
\begin{centering}
\includegraphics[bb=5bp 240bp 376bp 450bp,clip,width=1\columnwidth]{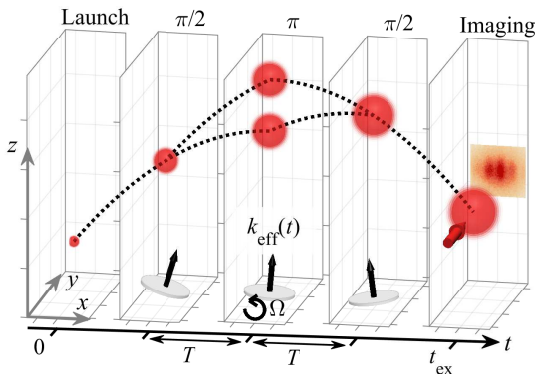}
\par\end{centering}
\centering{}\caption{Schematic diagram of point source atom interferometry. The atomic
cloud is launched upwards at $t=0$, interacts with a sequence of
$\pi/2$-$\pi$-$\pi/2$ two-photon Raman pulses comprising the interferometer,
and finally imaged at $t=t_{\textrm{ex}}$. The effective wave vector
$\mathbf{k}_{\textrm{eff}}$ of the counter-propagating Raman beams
is determined by a retro-reflecting mirror which can be tilted by
a piezo stage to simulate rotations. Ballistic expansion of the cloud
generates correlation between position and velocity of the atoms,
giving rise to a spatial interference pattern from which the rotation
rate is calculated. \label{fig:1-psi-scheme}}
\end{figure}

\section{Point-source atom interferometry}

PSI gyroscopes employs a Mach-Zehnder atom interferometer configuration
\citep{Kasevich_1991}. Three laser pulses, equally separated in time,
interact with a freely-falling atomic cloud and act, in analogy to
optical interferometry, as coherent beam splitters and mirrors. These
operations are realized through counter-propagating laser beams which
drive two-photon, Doppler-sensitive Raman transitions between different
internal ground state levels and different momentum states of the
atom \citep{Kasevich1991a}. Beam splitter operations correspond to
$\pi/2$-pulses which place the atom in a coherent superposition of
two momentum states separated by $\hbar\mathbf{k}_{\textrm{eff}}$,
where $\mathbf{k}_{\textrm{eff}}=\mathbf{k}_{1}-\mathbf{k}_{2}$ is
the two-photon wave vector of the Raman interaction. Likewise, mirror
operations correspond to $\pi$-pulses which change the momentum of
each interferometer arm by $\pm\hbar\mathbf{k}_{\textrm{eff}}$. During
the interferometer sequence, these pulses are utilized to coherently
split, redirect, and recombine the atomic wavepackets in space (Fig.\,\ref{fig:1-psi-scheme}).
The interferometer phase $\phi$ results from the different spatial
trajectories of the two arms, and determines the relative population
in the two internal states at the interferometer output. Through state-dependent
detection of the atoms, this relative population and thus the phase
can be measured.

In the Mach-Zehnder configuration considered here, the leading phase
contributions are given by $\phi_{\mathbf{a}}=\left(\mathbf{k}_{\textrm{eff}}T^{2}\right)\cdot\mathbf{a}$
and $\phi_{\boldsymbol{\Omega}}=\left(2\mathbf{v}\times\mathbf{k}_{\textrm{eff}}T^{2}\right)\cdot\boldsymbol{\Omega}$,
where $T$ is the time between each pair of pulses, $\mathbf{a}$
and $\boldsymbol{\Omega}$ are respectively the acceleration and rotation
rate of the atoms relative to the interrogating Raman beams, and $\mathbf{v}$
is the mean velocity of each atom during the interferometer sequence.
The acceleration phase $\phi_{\mathbf{a}}$ can be seen as resulting
from the space-time area enclosed by the two interferometer arms,
whereas the rotation phase $\phi_{\boldsymbol{\Omega}}$ results from
the enclosed spatial area, in a Sagnac-like effect.

\begin{figure}[t]
\begin{centering}
\includegraphics[bb=10bp 0bp 381bp 188bp,clip,width=1\columnwidth]{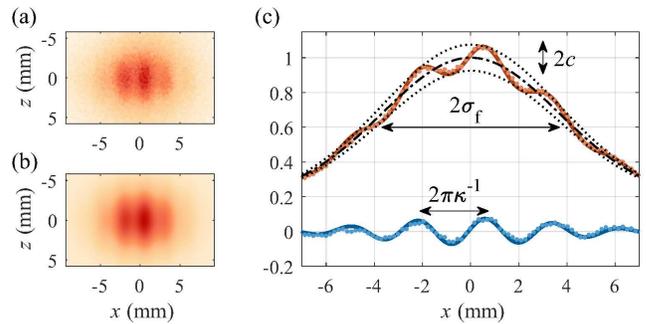}
\par\end{centering}
\centering{}\caption{Spatial fringe pattern measured with PSI. (a) Raw fluorescence image
from a single experiment. (b) Fit of the image to a two-dimensional
Gaussian distribution with sinusoidal modulation. (c) Normalized,
vertically integrated cross sections of the measured and fitted images
(red dotted and solid lines, respectively). The Gaussian density distribution
(dash-dotted line) of full width $2\sigma_{\textrm{f}}$ is modulated
by a fringe pattern (blue) with spatial frequency $\kappa$ and contrast
$c$. In this experiment, the initial cloud width was $2\sigma_{\textrm{i}}=\unit[0.98]{mm}$
(measured separately) with magnification of $\sigma_{\textrm{f}}/\sigma_{\textrm{i}}\sim7.5$.\label{fig:2-psi-images}}
\end{figure}

While $\phi_{\mathbf{a}}$ is uniform across the expanding atomic
cloud, $\phi_{\boldsymbol{\Omega}}$ depends on the initial transverse
velocity of each atom due to the cloud's finite temperature. In the
point-source limit, where the final cloud is much larger than its
initial size, ballistic expansion of the cloud results in one-to-one
correspondence between the atomic position and velocity. Thus, the
velocity-dependent rotation phase is projected onto a spatial fringe
pattern which may be directly imaged to facilitate a measurement of
$\boldsymbol{\Omega}$ \citep{Dickerson2013}.

In our setup, the Raman beams are aligned vertically such that $\mathbf{k}_{\textrm{eff}}=k_{\textrm{eff}}\hat{\mathbf{z}}$,
and the cloud is imaged in the $\hat{\mathbf{x}}\textrm{-}\hat{\mathbf{z}}$
plane, providing sensitivity to gravitational acceleration $-g\hat{\mathbf{z}}$
and to rotations around the $\mathbf{\hat{y}}$ axis. From here on,
we will focus on the latter term only. For simplicity we denote $\Omega=\Omega_{y}$
and write the rotation phase as $\phi_{\Omega}=2k_{\textrm{eff}}\Omega v_{x}T^{2}$.
In the point-source limit we have $x=v_{x}t_{\textrm{ex}}$, where
$t_{\textrm{ex}}$ is the total expansion time of the cloud, and the
rotation phase becomes $\phi_{\Omega}=\left(2k_{\textrm{eff}}\Omega T^{2}/t_{\textrm{ex}}\right)x$.
The resulting atomic density has a profile of a modulated Gaussian
distribution, as shown in Fig.\,\ref{fig:2-psi-images}, whose spatial
frequency is given by $\kappa=\left(2k_{\textrm{eff}}T^{2}/t_{\textrm{ex}}\right)\Omega$.
More generally we may write the spatial fringe frequency as $\kappa\left(\Omega\right)=F\Omega$
where $F$ is the scale factor, given in the point-source limit by
$F_{\textrm{ps}}=2k_{\textrm{eff}}T^{2}/t_{\textrm{ex}}$ \citep{Hoth2016}.

In general, realization of a rotation-sensing atom interferometry
requires separating the phase contributions of rotation and acceleration.
In PSI this is achieved by exploiting the velocity-dependence of the
rotation-sensitive phase $\phi_{\boldsymbol{\Omega}}$ to map it onto
a spatially-varying interference pattern, separating rotation and
acceleration into the spatial fringe frequency and phase, respectively
\citep{Chen2019}. A different approach utilizes two atomic sources
with opposite velocities, using either thermal \citep{Gustavson1997}
or cold atoms \citep{Canuel2006}, and performing identical Mach-Zehnder
sequences on both. Here too, the velocity-dependence of $\phi_{\boldsymbol{\Omega}}$
results in the output phases of the two interferometers to be $\phi_{\mathbf{a}}\pm\phi_{\boldsymbol{\Omega}}$,
allowing simple separation of the two contributions. Unlike PSI however,
in these schemes $\phi_{\boldsymbol{\Omega}}$ is uniform for each
of the two interferometers. Alternatively, a four-pulse ``butterfly''
interferometry sequence \citep{Savoie2018} may be used, which inherently
rejects the contribution of constant acceleration but remains sensitive
to time-varying acceleration, such as vibration noise. In comparison
to these two schemes, PSI offers a much simpler experimental configuration
using only a single atomic source and a single interrogation optical
beam, with suppression of both constant acceleration and vibrations.

\section{Experimental apparatus}

In our apparatus \citep{Yankelev2019,Avinadav2019}, we load an ensemble
of $^{87}\textrm{Rb}$ atoms from thermal background vapor in a magneto-optical
trap (MOT) and cool them to $\unit[4]{\mu K}$. Through moving optical
molasses, the atoms are launched vertically with velocities of up
to $\unit[1.2]{m/s}$ while occupying all Zeeman states $m_{F}=0,\pm1$
of the lower hyperfine level $F=1$. The interferometer pulses are
realized by counter-propagating Raman beams with $\sigma^{+}-\sigma^{+}$
polarizations that drive two-photon transitions between the $m_{F}=0$
states of $F=1$ and $F=2$, with a bias magnetic field of about $\unit[300]{mG}$
to separate the magnetic Zeeman states.

Raman beams are realized by phase modulation at $\sim\unit[6.834]{GHz}$
with an electro-optic modulator, with the carrier detuned $\unit[700]{MHz}$
red of the $F=2\rightarrow F'=1$ transition, followed by a two-stage
amplification to a total power of about $\unit[1]{W}$ in all sidebands.
The Raman beam is collimated to $\unit[70]{mm}$ diameter using a
Silicon Lightwave LB80 output collimator with $\lambda/10$ wavefront
error. The resulting two-photon Rabi frequency is $\unit[35]{kHz}.$

The retro-reflecting mirror of the Raman beams is mounted on an accurate
nanopositioning tip-tilt piezo stage (nPoint RXY3-410), which allows
us to rotate the optical wavevector during the interferometer sequence
and thereby mimic actual rotations. The stage has total dynamic range
of $\unit[\pm1.5]{mrad}$ in closed-loop operation, with $\unit[0.2]{\mu rad}$
position noise and a few milliseconds settling time.

Following the interferometer sequence, we use fluorescence excitation
on the optical $F=2\rightarrow F^{\prime}=3$ cycling transition,
employing all six MOT beams for minimal distortions, to capture an
image of the atoms occupying the $F=2$ level. From launch to detection,
the experimental sequence lasts in total up to $\unit[300]{ms}$.
The shot-to-shot cycle time is $\unit[3]{s}$ due to technical limitations
in communication bandwidth of the computer control system.

Images are taken on a PCO Pixelfly CCD camera, using a Fujinon-TV
H6X12.5R zoom lens. We fit the image from each experiment to a Gaussian
envelope function with sinusoidal modulation and extract $\kappa$,
from which $\Omega$ is inferred (Fig.\,\ref{fig:2-psi-images}).
We calibrate the imaging system by using velocity-selective Raman
pulses and their well-known timing and momentum transfer, to select
and image two narrow atomic ensembles at controllable distances.

\section{Finite-size effects}

\subsection{Theoretical model}

For clouds of finite initial size, the resulting atomic density profile
is a convolution of the initial density distribution with the ideal
point-source fringe pattern, which is a spatial fringe superimposed
on the unperturbed final density distribution. Mathematically, such
convolution modifies both the wavelength and contrast of the fringe
compared to the ideal point-source case. Hoth \emph{et al. }derived
analytically these effects assuming spherically-symmetric Gaussian
initial and final density distributions \citep{Hoth2017}. In this
case, both the spatial fringe frequency $\kappa$ and its contrast
$c$ are reduced with respect to their values in the point-source
limit. Defining the initial and final cloud widths as $\sigma_{\textrm{i}}$
and $\sigma_{\textrm{f}}$, respectively, the modified fringe properties
were found to be
\begin{align}
\kappa & =F_{\textrm{ps}}\Omega\left(1-\sigma_{\textrm{i}}^{2}/\sigma_{\textrm{f}}^{2}\right),\label{eq:k_Omega_finite}\\
c & =c_{0}\exp\left[-\frac{1}{2}\frac{\kappa^{2}\sigma_{\textrm{i}}^{2}}{1-\sigma_{\textrm{i}}^{2}/\sigma_{\textrm{f}}^{2}}\right],\label{eq:contrast_finite}
\end{align}
where $c_{0}$ is the interferometer contrast for $\Omega=0$.

In physical terms, imperfect correlation between the atoms' position
and velocity due to the finite-sized initial cloud results in averaging
over atoms with different velocities at each measured position. For
a Gaussian velocity distribution, the average velocity of atoms at
each position is smaller than the expected value in the point-source
limit, resulting in a reduction of the spatial fringe frequency. Similarly,
the average over different velocities smears the interference fringes
and reduces their contrast.

As a consequence of Eq.\,(\ref{eq:k_Omega_finite}), the scale factor
of the interferometer becomes $F=F_{\textrm{ps}}(1-\sigma_{\textrm{i}}^{2}/\sigma_{\textrm{f}}^{2})$.
The dependency on $\sigma_{\textrm{i}}/\sigma_{\textrm{f}}$ leads
to bias instability of the sensor when either $\sigma_{\textrm{i}}$
or $\sigma_{\textrm{f}}$ drift. Importantly, for a given relative
drift in $\sigma_{\textrm{i}}$, the drift in the scale factor increases
with the inverse magnification ratio $\sigma_{\textrm{i}}/\sigma_{\textrm{f}}$,
emphasizing the susceptibility of compact sensors with small magnification
to such drifts.

A larger magnification ratio can be achieved by decreasing $\sigma_{\textrm{i}}$
using, \emph{e.g.}, tight dipole traps, which adds experimental complexity
and may impact the atom number; or by increasing $\sigma_{\textrm{f}}$
through longer expansion times or higher cloud temperatures, in either
case requiring larger atom optics and detection beams. However, even
when operating at a large magnification ratio such as $\sigma_{\textrm{f}}/\sigma_{\textrm{i}}=50$,
the scale factor is modified by $\unit[400]{ppm}$ and small drifts
in the initial cloud size on the order of a few percent would result
in changes of tens parts per million to the scale factor. In comparison,
optical gyroscopes achieve scale factor stability on the order of
$\unit[1]{ppm}$ \citep{Lefevre2014a}.

\subsection{Experimental verification}

\begin{figure}[b]
\begin{centering}
\includegraphics[bb=5bp 0bp 376bp 300bp,clip,width=1\columnwidth]{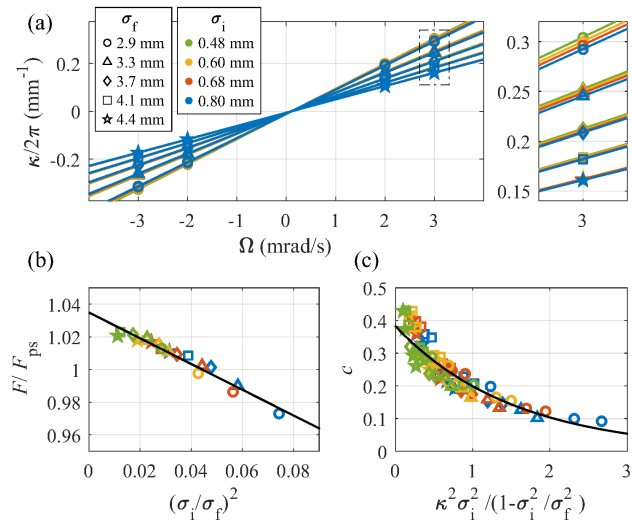}
\par\end{centering}
\centering{}\caption{Effect of initial cloud size $\sigma_{\textrm{i}}$ and final cloud
size $\sigma_{\textrm{f}}$ on PSI scale factor and contrast. (a)
Measured spatial fringe frequency $\kappa$ as a function of $\Omega$
for different values of $\sigma_{\textrm{i}}$ and $\sigma_{\textrm{f}}$.
Solid lines are linear fits, with slopes representing the scale factors.
Different values of $\sigma_{\textrm{f}}$ correspond to launch velocities
$\unit[0.6\textrm{-}1.2]{m/s}$, with $t_{\textrm{ex}}$ between $\unit[133\textrm{-}252]{ms}$.
In all cases $T=\unit[52]{ms}$. (b) Measured scale factor for different
values of $\sigma_{\textrm{i}}$ and $\sigma_{\textrm{f}}$. Solid
line is a linear fit with slope $-0.79(4)$, deviating from the predicted
value $-1$. (c) Measured fringe contrast for different values of
$\sigma_{\textrm{i}}$ and $\sigma_{\textrm{f}}$. Solid line is an
exponential fit to $c=c_{0}\exp\left[-\alpha\kappa^{2}\sigma_{\textrm{i}}^{2}/\left(1-\sigma_{\textrm{i}}^{2}/\sigma_{\textrm{f}}^{2}\right)\right]$
with $\alpha=0.65(3)$, deviating from the model prediction $\alpha=0.5$.\label{fig:3-finite-source-effects}}
\end{figure}

We verify the model predictions by performing measurements with different
values of $\sigma_{\textrm{i}}$, obtained through different repump
intensities at the final optical molasses stage, and with different
values $\sigma_{\textrm{f}}$, obtained through different launch velocities
and thus different expansion times. $\sigma_{\textrm{i}}$ was measured
in independent experiments by imaging the cloud immediately after
launch. For standard operation of the optical molasses, the value
of $\sigma_{\textrm{i}}$ is about $\unit[0.5]{mm}$, and we can increase
it up to about $\unit[\ensuremath{0.8}]{mm}$ before suffering significant
loss of atoms. As Fig.\,\ref{fig:3-finite-source-effects} shows,
we find good agreement with the functionals in Eqs.\,(\ref{eq:k_Omega_finite})
and (\ref{eq:contrast_finite}) for a range of magnification ratios
$\sigma_{\textrm{f}}/\sigma_{\textrm{i}}$ between $3.5$ and $9$,
corresponding to changes of 5\% in scale factor. The fact that the
scale factor $F$ at $\sigma_{\textrm{i}}/\sigma_{\textrm{f}}\rightarrow0$
differs from $F_{\textrm{ps}}$ by 3.5\% can be attributed to imaging
distortions and to residual calibration errors of the imaging system
and of the piezo stage driving the mirror rotations. The fitted slope
for the scale-factor dependence on $\sigma_{\textrm{i}}^{2}/\sigma_{\textrm{f}}^{2}$,
as well as the scaling of the exponential decrease of the contrast,
slightly differ from the theoretical model either due to estimation
errors in $\sigma_{\textrm{i}}$ and $\sigma_{\textrm{f}}$ or due
to non-Gaussian density distributions.

\section{Contrast-based stabilization}

\subsection{Model description}

The first method we present for correcting scale factor instability
exploits the correlation between $\Omega$ and $\left(\kappa,c,\sigma_{\textrm{f}}\right)$
which are extracted from each PSI image. We obtain the relation $\Omega\left(\kappa,c,\sigma_{\textrm{f}}\right)$
by eliminating $\sigma_{\textrm{i}}$ from Eqs.\,(\ref{eq:k_Omega_finite})
and (\ref{eq:contrast_finite}). To capture the aforementioned deviations
from the model, we add a single parameter $\beta$ and replace $\sigma_{\textrm{f}}\rightarrow\sigma_{\textrm{f}}/\beta$.
The resulting expression for $\Omega$ is then
\begin{equation}
\Omega=\frac{\kappa}{F_{\textrm{ps}}}\left[1-2\left(\frac{\beta}{\kappa\sigma_{\textrm{f}}}\right)^{2}\ln\frac{c}{c_{0}}\right].\label{eq:k_ps_self_correction}
\end{equation}
As demonstrated below, we verify that this formulation, based on one
physical parameter $c_{0}$ and one correction parameter $\beta$,
provides an excellent description of the measured data. Importantly,
it allows us to calculate $\Omega$ using only the parameters $\kappa$,
$c$ and $\sigma_{\textrm{f}}$ measured in every single PSI image,
enabling analysis on a single-shot basis without any reduction in
temporal bandwidth.

\subsection{Parameter calibration}

\begin{figure}[t]
\begin{centering}
\includegraphics[bb=5bp 0bp 386bp 263bp,clip,width=1\columnwidth]{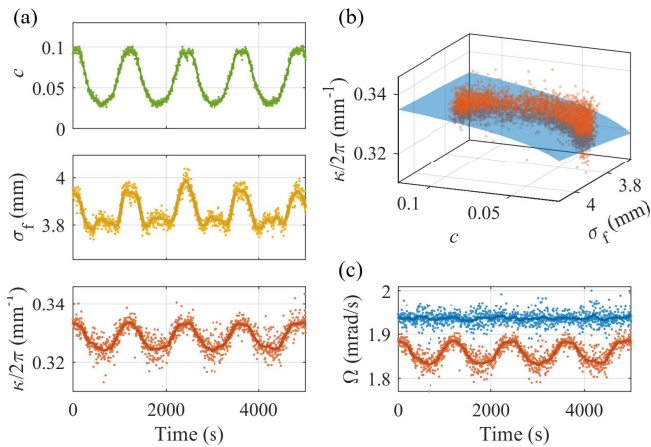}
\par\end{centering}
\centering{}\caption{Calibration of contrast-based scale-factor stabilization. (a) Top
to bottom: measured contrast, final cloud size, and spatial fringe
frequency for an applied rotation rate of $\unit[2]{mrad/s}$, as
$\sigma_{\textrm{i}}$ is varied periodically in the range $\unit[0.5-0.8]{mm}$.
Each data point represents a single PSI image. In these and following
experiments, $T=\unit[80]{ms}$ and $t_{\textrm{ex}}=\unit[185]{ms}$.
(b) Parametric plot of $\kappa\left(c,\sigma_{\textrm{f}}\right)$.
Dots are data points, surface is a fit to Eq.\,(\ref{eq:k_calib_curve})
with $c_{0}=0.43(8)$ and $\beta=0.80(1)$. The fit describes the
data points well, with residuals of $\kappa$ from the fit surface
showing no correlation to $c$ or $\sigma_{\textrm{f}}$ and characterized
by standard deviation $\unit[\left(2\pi\right)2.49\times10^{-3}]{mm^{-1}}$,
consistent with a moving standard deviation $\unit[\left(2\pi\right)2.55\times10^{-3}]{mm^{-1}}$
of the raw $\kappa$ measurements. (c) Rotation rate calculated with
the uncorrected scale-factor $F_{\textrm{ps}}$ (red), compared to
corrected rotation rate (blue), which employs Eq.\,(\ref{eq:k_ps_self_correction})
with the calibrated parameters, showing excellent rejection of scale-factor
drifts. Solid lines in (a),(c) are 30-samples moving averages.\label{fig:4-contrast-calibration}}
\end{figure}

To calibrate $c_{0}$ and $\beta$ in Eq.\,(\ref{eq:k_ps_self_correction}),
we perform measurements at a constant rotation rate and periodically
vary $\sigma_{\textrm{i}}$ by changing the repump beam intensity
during the moving optical molasses. This results in changes to $c$,
$\sigma_{\textrm{f}}$, and $\kappa$, as shown in Fig.\,\ref{fig:4-contrast-calibration}(a).
Inverting Eq.\,(\ref{eq:k_ps_self_correction}) and setting $\Omega=\Omega_{\textrm{calib}}$,
we obtain an expression for $\kappa$ in this calibration measurement,

\begin{equation}
\kappa=\frac{1}{2}F_{\textrm{ps}}\Omega_{\textrm{calib}}\left[1+\sqrt{1+\frac{8\ln\left(c/c_{0}\right)}{\left(F_{\textrm{ps}}\Omega_{\textrm{calib}}\sigma_{\textrm{f}}/\beta\right)^{2}}}\right].\label{eq:k_calib_curve}
\end{equation}
The measured data points are fitted to this surface equation, as shown
in Fig.\,\ref{fig:4-contrast-calibration}(b), where $c_{0}$, $\beta$,
and $\Omega_{\textrm{calib}}$ are the fit parameters, the latter
being an auxiliary parameter not used in any subsequent analysis.

With calibrated $c_{0}$ and $\beta$ at hand, we now use Eq.\,(\ref{eq:k_ps_self_correction})
to correct the inferred rotation rate in this calibration run. This
exercise is shown in Fig.\,\ref{fig:4-contrast-calibration}(c),
and indeed we find that the drifts are removed and a stable rotation
measurement is obtained. We emphasize that the analysis does not use
any assumption or prior knowledge on $\sigma_{\textrm{i}}$ or on
its temporal variations, and in fact the only prerequisite is that
the range of $\sigma_{\textrm{i}}$ scanned during the calibration
process is large enough to adequately calibrate the model parameters.

While this correction is aimed to improve the measurement stability
under variations of $\sigma_{\textrm{i}}$, the measurement accuracy
is enhanced as well. As Eq.\,(\ref{eq:k_ps_self_correction}) captures
also the systematic bias due to the finite magnification ratio, as
described by Eq.\,(\ref{eq:k_Omega_finite}), applying this correction
suppresses the systematic bias. Indeed, we find the post-correction
estimated rotation rate {[}Fig.\,\ref{fig:4-contrast-calibration}(c){]}
to be $\unit[1.936(6)]{mrad/s}$, in agreement with the expected value
$\unit[1.9387]{mrad/s}$ which consists of the applied mirror rotation
rate of $\unit[2]{mrad/s}$ and the projection of Earth's rotation
on the measurement axis, which was due north, at our latitude of $\unit[32.7940^{\circ}]{N}$.
In contrast, the mean estimated rotation rate using the point-source
scale factor is $\unit[1.857]{mrad/s}$, differing significantly from
the true value.

Nevertheless, it is possible that some residual bias still exists
after applying Eq.\,(\ref{eq:k_ps_self_correction}). However, it
would be inseparable from other possible sources of systematic bias
in the PSI measurement due to, \emph{e.g.}, effect of wavefront aberrations
\citep{Gauguet2009,Schkolnik2015,Karcher2018} or imaging distortions.
Calibration error of the piezo stage and inaccurate estimation of
the measurement axis north alignment will contribute to an apparent
bias as well. For this reason $\Omega_{\textrm{calib}}$ is treated
as a fit parameter when using Eq.\,(\ref{eq:k_calib_curve}), rather
than being simply taken as the input rotation rate. The overall bias
may be corrected by calibrating the PSI sensor, for example using
a precision rotary stage.

\subsection{Demonstration of correction performance}

\begin{figure}[b]
\begin{centering}
\includegraphics[bb=5bp 0bp 376bp 375bp,clip,width=1\columnwidth]{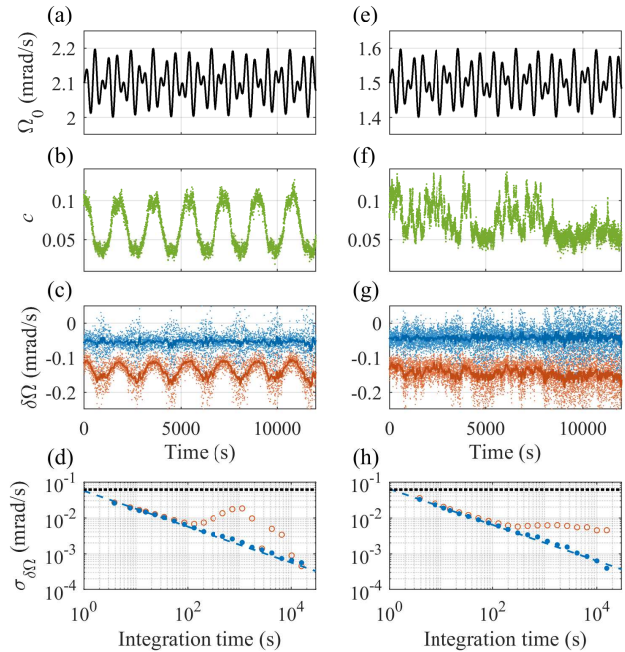}
\par\end{centering}
\centering{}\caption{Demonstration of contrast-based scale-factor stabilization. (a) Time
series of the input rotation rate $\Omega_{0}$, oscillating around
$\unit[2.1]{mrad/s}$. (b) Measured fringe contrast shows variations
due to applied periodic changes to $\sigma_{\textrm{i}}$. (c) Residuals
of the estimated rotation rate $\Omega-\Omega_{0}$, with (blue) and
without (red) scale-factor correction using Eq.\,(\ref{eq:k_ps_self_correction}).
Solid lines are 30-samples moving averages. The full data spans $\unit[3\times10^{4}]{sec}$.
(d) Allan deviation of the residuals, with and without scale factor
correction. Dashed line is a fit to $\tau^{-1/2}$. Dotted line indicates
Earth rotation rate at our lab location. (e-h) Same as (a-d), with
$\Omega_{0}$ oscillating around $\unit[1.5]{mrad/s}$ and random-walk
variations in $\sigma_{\textrm{i}}$ in the range $\unit[0.5-0.8]{mm}$.
The observed sensitivities are $\unit[28]{\mu rad/s}$, $\unit[32]{\mu rad/s}$
per shot in panels (d), (h), respectively, differing slightly due
to changing experimental conditions, such as total atom number.\label{fig:5-contrast-stabilization}}
\end{figure}

We demonstrate the implementation of this stabilization method in
two separate measurement runs, as presented in Fig.\,\ref{fig:5-contrast-stabilization},
representing realistic operating scenarios where the rotation rate
varies in time. We artificially generate scale-factor drifts by varying
the optical molasses parameters which affect the initial and final
cloud distributions, both in periodic fashion and in random-walk-like
behavior. While the rotation rate and initial conditions change smoothly
in time, this information is not used in the analysis as all corrections
are on a single-shot basis. The interferometer parameters are $T=\unit[80]{ms}$
and $t_{\textrm{ex}}=\unit[185]{ms}$, as in the calibration run.

Using Eq.\,(\ref{eq:k_ps_self_correction}) and the previously calibrated
values for $c_{0}$ and $\beta$, we obtain suppression of sensor
drifts by a factor of ten in both examples {[}Fig.\,\ref{fig:5-contrast-stabilization}(d),(h){]},
limited only by the magnitude of the drifts we introduced. The scheme
allows complete recovery of $\tau^{-1/2}$ averaging performance at
time scales up to $10^{4}$ seconds, affirming that it does not introduce
new bias instabilities of its own. The second example, where $\Omega$
varies around a mean value smaller than $\Omega_{\textrm{calib}}$,
and the model measurements shown in Fig. \ref{fig:3-finite-source-effects}
with larger values of $\Omega$, demonstrate that the technique is
robust and does not require multiple calibration runs with different
values of $\Omega_{\textrm{calib}}$. Our measurements reach a stability
of $\unit[0.5]{\mu rad/s}$, providing a upper bound for the scale-factor
stability of $2.5\times10^{-4}$. Under the applied variation of $\pm18\%$
in $\sigma_{\textrm{i}}$ , reaching such stability without any correction
scheme would require operating the PSI sensor at a magnification ratio
of at least $40$.

Finally, we note that it is possible in principle to apply a correction
similar to that described in this section by utilizing Eq.\,(\ref{eq:k_Omega_finite})
and directly measuring $\sigma_{\textrm{i}}$. This may be done either
in separate experimental shots, at the expense of reduced bandwidth
and sensitivity per $\sqrt{\textrm{Hz}}$ and under the constraint
that $\sigma_{\textrm{i}}$ changes little between shots, or through
non-destructive imaging of the cloud during the final moving molasses
stage. The latter is challenging as it requires short exposure time
to avoid motion blur of the small, moving cloud. In addition to these
constraints, as Fig.\,\ref{fig:3-finite-source-effects}(b) demonstrates,
Eq.\,(\ref{eq:k_Omega_finite}) requires an additional scaling parameter
to fit the measured data, likely due to the non-Gaussian shape of
the initial cloud. Thus this approach of directly measuring $\sigma_{\textrm{i}}$
does not eliminate the need for calibration parameters.

\section{Self-calibrating stabilization}

\subsection{Principles and demonstration}

Our second method for scale-factor stabilization utilizes the piezo
rotation stage to apply a known rotation $\Omega_{\textrm{bias}}$
at alternating measurements, in addition to the unknown rotation $\Omega$
which we wish to measure. Denoting the scale factor in these measurements
as $F$, the spatial fringe frequencies of two consecutive measurements
are given by $\kappa_{1}=F\Omega$ and $\kappa_{2}=F\left(\Omega+\Omega_{\textrm{bias}}\right)$.
These equations may be inverted to yield
\begin{align}
\Omega & =\Omega_{\textrm{bias}}\frac{\kappa_{1}}{\kappa_{2}-\kappa_{1}},\label{eq:rotation_modulation_Omega}\\
F & =\frac{1}{\Omega_{\textrm{bias}}}\left(\kappa_{2}-\kappa_{1}\right),\label{eq:rotation_modulation_scale_factor}
\end{align}
allowing us to extract both $\Omega$ and $F$ from each pair of measurements
$\kappa_{1}$ and $\kappa_{2}$. This self-calibrating method requires
no prior knowledge or assumption of a model describing the relationship
between $F$ and other system parameters. The signs of $\kappa_{1},\kappa_{2}$
are not directly available from the PSI images but rather assumed
to be known, for example from an auxiliary rotation sensor, as in
all existing schemes of PSI.

\begin{figure}[t]
\begin{centering}
\includegraphics[bb=5bp 0bp 386bp 338bp,clip,width=1\columnwidth]{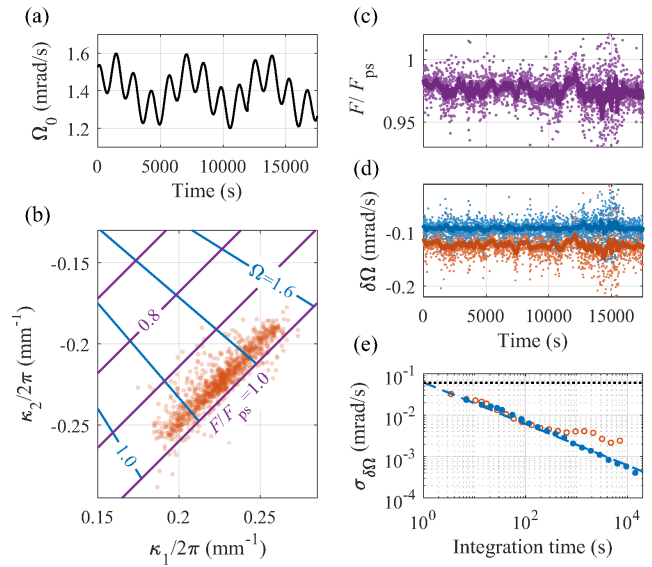}
\par\end{centering}
\centering{}\caption{Demonstration of scale-factor stabilization by self-calibration. (a)
Time series of the input rotation rate $\Omega_{0}$. (b) Measured
$\kappa_{2}$ and $\kappa_{1}$, each dot corresponding to a pair
of experiments with and without $\Omega_{\textrm{bias}}=\unit[-2.6]{mrad/s}$.
Also shown are contour lines of the estimated rotation rate (blue)
and scale factor correction relative to $F_{\textrm{ps}}$ (purple)
based on Eqs.\,(\ref{eq:rotation_modulation_Omega}-\ref{eq:rotation_modulation_scale_factor}).
(c) Estimated scale factor correction from each pair of measurements,
showing random-walk variations due to applied changes to $\sigma_{\textrm{i}}$
in the range $\unit[0.5-0.8]{mm}$. (d) Residuals of the estimated
rotation rate $\Omega-\Omega_{0}$, using the nominal scale factor
$F=F_{\textrm{ps}}$ (red) and the corrected scale factor (blue).
Solid lines in (c),(d) are 30-samples moving averages. (e) Allan deviation
of the residuals, with and without scale factor correction. Dashed
line is a fit to $\tau^{-1/2}$ with sensitivity $\unit[32]{\mu rad/s}$
per shot.\label{fig:6-rotation-modulation}}
\end{figure}

An experimental demonstration of the self-calibration method is shown
in Fig.\,\ref{fig:6-rotation-modulation}, with both the measured
rotation $\Omega$ and the scale factor changing in time. Again we
find perfect suppression of scale factor drifts and recovery of $\tau^{-1/2}$
noise behavior. Compared to the previously described contrast-based
stabilization, this method reduces the effective bandwidth by a factor
of two due to the pairwise analysis but has the advantages of being
model-independent and not requiring any initial calibration measurements.
We note that despite the reduction in bandwidth, information is not
lost when analyzing pairs of measurements and thus the rotation sensitivity
per $\sqrt{\textrm{Hz}}$ is maintained, as evident in Fig.\,\ref{fig:6-rotation-modulation}(e).

This correction scheme relies on an accurate piezo stage for adding
a known bias rotation, as errors in $\Omega_{\textrm{bias}}$ enter
directly into the estimation of $\Omega$ itself. Therefore the stage
is required to have similar resolution as the PSI measurement so as
not to degrade its sensitivity, and long-term stability to enable
scale-factor corrections at long times. As our results show, these
conditions are fulfilled in our apparatus. Indeed, such a stage with
adequate performance is likely to be necessary anyway in an operational
rotation sensor, for purposes such as calibration and dynamic range
compensation.

\subsection{Generalization to time-varying scenarios}

The analysis above assumed that $\Omega$ and $F$ are constant in
both measurements. We now turn to the more general case where these
quantities may change between shots. We define $\Omega_{1}=\Omega+\delta\Omega$,
$\Omega_{2}=\Omega-\delta\Omega+\Omega_{\textrm{bias}}$, and $F_{1,2}=F_{0}+\delta F_{\textrm{comm}}\pm\delta F_{\textrm{diff}}$,
such that the measured spatial fringe frequencies are $\kappa_{1,2}=F_{1,2}\Omega_{1,2}$.
Here $\delta\Omega$ represents shot-to-shot variation in $\Omega$,
$F_{0}$ is the nominal scale-factor, and $\delta F_{\textrm{comm}}$
and $\delta F_{\textrm{diff}}$ are scale-factor errors common and
different, respectively, in both measurements. $\Omega_{\textrm{bias}}$
as before is the known rotation which is added in the second measurement.
Using Eq.\,(\ref{eq:rotation_modulation_Omega}) and to leading order
in $\delta\Omega/\Omega$ and $\delta F/F$, we find that the estimated
rotation rate is
\begin{align}
\frac{\Omega_{\textrm{est}}}{\Omega} & \approx1+\left(2+\frac{2\Omega}{\Omega_{\textrm{bias}}}\right)\frac{\delta F_{\textrm{diff}}}{F_{0}}+\left(1+\frac{2\Omega}{\Omega_{\textrm{bias}}}\right)\frac{\delta\Omega}{\Omega}\label{eq:Omega_self_calib_est}
\end{align}
To eliminate the $\delta\Omega$-related term in Eq.\,(\ref{eq:Omega_self_calib_est}),
we set $\Omega_{\textrm{bias}}=-2\left(\kappa_{1}/F_{0}\right)$,
where $\kappa_{1}/F_{0}$ corresponds to an estimate of $\Omega_{1}$
based on the first measurement and assuming the nominal scale-factor.
With this choice of $\Omega_{\textrm{bias}}$, we have to leading
order $\Omega_{\textrm{est}}\approx\Omega\left(1+\delta F_{\textrm{diff}}/F_{0}\right)$,
such that the estimated rotation rate corresponds to the average $\Omega$
of the two measurements.

Regardless of the exact choice of $\Omega_{\textrm{bias}}$, we find
that even in this case of varying-rotation rate and scale-factor,
the result is insensitive to errors in the scale-factor which are
common to both measurements and remains sensitive only to non-common
scale-factor errors $\delta F_{\textrm{diff}}$. In contrast, the
``naive'' measurement approach, with $\Omega_{\textrm{bias}}=0$
and estimating $\Omega$ as the average $\frac{1}{2}\left(\kappa_{1}+\kappa_{2}\right)/F_{0}$,
leads to $\Omega_{\textrm{naive}}\approx\Omega\left(1+\delta F_{\textrm{comm}}/F_{0}\right)$.
It is then the non-common scale-factor error which cancels to leading
order, and the common error remains.

In most scenarios, $\delta F_{\textrm{comm}}\gg\delta F_{\textrm{diff}}$
\emph{i.e.}, the scale-factor is expected to change slowly with respect
to the time scale of two measurements, and thus our self-calibrating
method cancels the more dominant source of error. Additionally, more
advanced estimation protocols such as a Kalman filter \citep{Kalman1960,Cheiney2018}
could be used to track rapidly varying rotation signals while estimating
the slowly-varying scale factor from multiple measurement pairs.

\section{Conclusion}

We have presented two complementary approaches for improving the stability
of rotation measurements using point source atom interferometry, suppressing
scale factor drifts which arise due to changes in the atomic cloud
parameters. We demonstrated the two schemes experimentally with complete
suppression of scale factor drifts. We showed that they maintain the
sensor sensitivity and do not introduce any drifts on their own. We
achieved scale-factor stability on time scales up to $\unit[10^{4}]{sec}$,
representing orders of magnitude improvement in stability time compared
to previous works. We reach rotation rate stability of $\unit[0.5]{\mu rad/s}$.

The first approach utilizes the inherent correlations between different
parameters in PSI images, namely the contrast, spatial frequency and
final size of the fringe pattern, for estimating the scale factor
from a single image and correct for drifts. The underlying model which
describes these correlations is based on the physics of the sensor,
rather than on an empirical correlation, as we verify by independent
measurements. This method is based on information which is already
available in standard PSI measurements and maintains the original
bandwidth of the sensor, however preliminary calibration of two model
parameters is necessary. The possibility of using a continuous estimation
protocol such as Kalman or particle filtering \citep{Moral1997},
to replace the initial calibration as well as to allow temporal variations
of these parameters, may also be explored.

The second approach relies on sequential measurements with an added
bias rotation to directly estimate the scale factor from each pair
of measurements. This approach is self-calibrating and completely
model-independent, but it depends on the stability of the piezo mirror
stage generating the bias rotation and reduces the sensor bandwidth.

While the experiments described here focus on side-imaging PSI which
allows single-axis rotation sensing, the two schemes we developed
are fully compatible under the same conditions with dual-axis sensors
using top- or bottom- imaging of the cloud.

The stabilization techniques we developed are particularly important
in compact PSI sensors, which are an attractive alternative for constructing
simple, mobile device with high sensitivity to rotations. Our schemes
address a significant challenge of such PSI sensors, namely high susceptibility
to scale factor drifts due to the low magnification ratio of the atomic
cloud. The results may pave the way for realizing high-performance,
compact PSI devices for demanding applications that require long-time
stability, such as gyrocompasses for north finding applications, gyroscopes
for line-of-sight stabilization, or inertial measurement units for
navigation systems.

\section*{Acknowledgments}

This work was supported by the Pazy Foundation, the Israel Science
Foundation and the Quantum Technologies Development Consortium (RA).
We thank Shlomi Kotler for fruitful discussions.

\bibliography{PSI_Scale_factor_stabilization}

\begin{thebibliography}{44}%
\makeatletter
\providecommand \@ifxundefined [1]{%
 \@ifx{#1\undefined}
}%
\providecommand \@ifnum [1]{%
 \ifnum #1\expandafter \@firstoftwo
 \else \expandafter \@secondoftwo
 \fi
}%
\providecommand \@ifx [1]{%
 \ifx #1\expandafter \@firstoftwo
 \else \expandafter \@secondoftwo
 \fi
}%
\providecommand \natexlab [1]{#1}%
\providecommand \enquote  [1]{``#1''}%
\providecommand \bibnamefont  [1]{#1}%
\providecommand \bibfnamefont [1]{#1}%
\providecommand \citenamefont [1]{#1}%
\providecommand \href@noop [0]{\@secondoftwo}%
\providecommand \href [0]{\begingroup \@sanitize@url \@href}%
\providecommand \@href[1]{\@@startlink{#1}\@@href}%
\providecommand \@@href[1]{\endgroup#1\@@endlink}%
\providecommand \@sanitize@url [0]{\catcode `\\12\catcode `\$12\catcode
  `\&12\catcode `\#12\catcode `\^12\catcode `\_12\catcode `\%12\relax}%
\providecommand \@@startlink[1]{}%
\providecommand \@@endlink[0]{}%
\providecommand \url  [0]{\begingroup\@sanitize@url \@url }%
\providecommand \@url [1]{\endgroup\@href {#1}{\urlprefix }}%
\providecommand \urlprefix  [0]{URL }%
\providecommand \Eprint [0]{\href }%
\providecommand \doibase [0]{http://dx.doi.org/}%
\providecommand \selectlanguage [0]{\@gobble}%
\providecommand \bibinfo  [0]{\@secondoftwo}%
\providecommand \bibfield  [0]{\@secondoftwo}%
\providecommand \translation [1]{[#1]}%
\providecommand \BibitemOpen [0]{}%
\providecommand \bibitemStop [0]{}%
\providecommand \bibitemNoStop [0]{.\EOS\space}%
\providecommand \EOS [0]{\spacefactor3000\relax}%
\providecommand \BibitemShut  [1]{\csname bibitem#1\endcsname}%
\let\auto@bib@innerbib\@empty
\bibitem [{\citenamefont {Tino}\ and\ \citenamefont
  {Kasevich}(2014)}]{Tino2014}%
  \BibitemOpen
  \bibinfo {editor} {\bibfnamefont {G.~M.}\ \bibnamefont {Tino}}\ and\ \bibinfo
  {editor} {\bibfnamefont {M.~A.}\ \bibnamefont {Kasevich}},\ eds.,\ \href@noop
  {} {\emph {\bibinfo {title} {Atom Interferometry, in Proceedings of the
  International School of Physics "Enrico Fermi," Course CLXXXVIII}}}\
  (\bibinfo  {publisher} {Societa Italiana di Fisica and IOS Press},\ \bibinfo
  {year} {2014})\BibitemShut {NoStop}%
\bibitem [{\citenamefont {Weiss}\ \emph {et~al.}(1993)\citenamefont {Weiss},
  \citenamefont {Young},\ and\ \citenamefont {Chu}}]{Weiss1993}%
  \BibitemOpen
  \bibfield  {author} {\bibinfo {author} {\bibfnamefont {D.~S.}\ \bibnamefont
  {Weiss}}, \bibinfo {author} {\bibfnamefont {B.~C.}\ \bibnamefont {Young}}, \
  and\ \bibinfo {author} {\bibfnamefont {S.}~\bibnamefont {Chu}},\ }\bibfield
  {title} {\enquote {\bibinfo {title} {Precision measurement of the photon
  recoil of an atom using atomic interferometry},}\ }\href {\doibase
  10.1103/physrevlett.70.2706} {\bibfield  {journal} {\bibinfo  {journal}
  {Physical Review Letters}\ }\textbf {\bibinfo {volume} {70}} (\bibinfo {year}
  {1993}),\ 10.1103/physrevlett.70.2706}\BibitemShut {NoStop}%
\bibitem [{\citenamefont {Dimopoulos}\ \emph {et~al.}(2007)\citenamefont
  {Dimopoulos}, \citenamefont {Graham}, \citenamefont {Hogan},\ and\
  \citenamefont {Kasevich}}]{Dimopoulos2007}%
  \BibitemOpen
  \bibfield  {author} {\bibinfo {author} {\bibfnamefont {S.}~\bibnamefont
  {Dimopoulos}}, \bibinfo {author} {\bibfnamefont {P.~W.}\ \bibnamefont
  {Graham}}, \bibinfo {author} {\bibfnamefont {J.~M.}\ \bibnamefont {Hogan}}, \
  and\ \bibinfo {author} {\bibfnamefont {M.~A.}\ \bibnamefont {Kasevich}},\
  }\bibfield  {title} {\enquote {\bibinfo {title} {Testing general relativity
  with atom interferometry},}\ }\href {\doibase 10.1103/physrevlett.98.111102}
  {\bibfield  {journal} {\bibinfo  {journal} {Physical Review Letters}\
  }\textbf {\bibinfo {volume} {98}} (\bibinfo {year} {2007}),\
  10.1103/physrevlett.98.111102}\BibitemShut {NoStop}%
\bibitem [{\citenamefont {M\"{u}ller}\ \emph {et~al.}(2010)\citenamefont
  {M\"{u}ller}, \citenamefont {Peters},\ and\ \citenamefont
  {Chu}}]{Mueller2010}%
  \BibitemOpen
  \bibfield  {author} {\bibinfo {author} {\bibfnamefont {H.}~\bibnamefont
  {M\"{u}ller}}, \bibinfo {author} {\bibfnamefont {A.}~\bibnamefont {Peters}},
  \ and\ \bibinfo {author} {\bibfnamefont {S.}~\bibnamefont {Chu}},\ }\bibfield
   {title} {\enquote {\bibinfo {title} {A precision measurement of the
  gravitational redshift by the interference of matter waves},}\ }\href
  {\doibase 10.1038/nature08776} {\bibfield  {journal} {\bibinfo  {journal}
  {Nature}\ }\textbf {\bibinfo {volume} {463}} (\bibinfo {year} {2010}),\
  10.1038/nature08776}\BibitemShut {NoStop}%
\bibitem [{\citenamefont {Bouchendira}\ \emph {et~al.}(2011)\citenamefont
  {Bouchendira}, \citenamefont {Clad{\'{e}}}, \citenamefont
  {Guellati-Kh{\'{e}}lifa}, \citenamefont {Nez},\ and\ \citenamefont
  {Biraben}}]{Bouchendira2011}%
  \BibitemOpen
  \bibfield  {author} {\bibinfo {author} {\bibfnamefont {R.}~\bibnamefont
  {Bouchendira}}, \bibinfo {author} {\bibfnamefont {P.}~\bibnamefont
  {Clad{\'{e}}}}, \bibinfo {author} {\bibfnamefont {S.}~\bibnamefont
  {Guellati-Kh{\'{e}}lifa}}, \bibinfo {author} {\bibfnamefont {F.}~\bibnamefont
  {Nez}}, \ and\ \bibinfo {author} {\bibfnamefont {F.}~\bibnamefont
  {Biraben}},\ }\bibfield  {title} {\enquote {\bibinfo {title} {New
  determination of the fine structure constant and test of the quantum
  electrodynamics},}\ }\href {\doibase 10.1103/physrevlett.106.080801}
  {\bibfield  {journal} {\bibinfo  {journal} {Physical Review Letters}\
  }\textbf {\bibinfo {volume} {106}} (\bibinfo {year} {2011}),\
  10.1103/physrevlett.106.080801}\BibitemShut {NoStop}%
\bibitem [{\citenamefont {Rosi}\ \emph {et~al.}(2014)\citenamefont {Rosi},
  \citenamefont {Sorrentino}, \citenamefont {Cacciapuoti}, \citenamefont
  {Prevedelli},\ and\ \citenamefont {Tino}}]{Rosi2014}%
  \BibitemOpen
  \bibfield  {author} {\bibinfo {author} {\bibfnamefont {G.}~\bibnamefont
  {Rosi}}, \bibinfo {author} {\bibfnamefont {F.}~\bibnamefont {Sorrentino}},
  \bibinfo {author} {\bibfnamefont {L.}~\bibnamefont {Cacciapuoti}}, \bibinfo
  {author} {\bibfnamefont {M.}~\bibnamefont {Prevedelli}}, \ and\ \bibinfo
  {author} {\bibfnamefont {G.~M.}\ \bibnamefont {Tino}},\ }\bibfield  {title}
  {\enquote {\bibinfo {title} {Precision measurement of the newtonian
  gravitational constant using cold atoms},}\ }\href {\doibase
  10.1038/nature13433} {\bibfield  {journal} {\bibinfo  {journal} {Nature}\
  }\textbf {\bibinfo {volume} {510}} (\bibinfo {year} {2014}),\
  10.1038/nature13433}\BibitemShut {NoStop}%
\bibitem [{\citenamefont {Zhou}\ \emph {et~al.}(2015)\citenamefont {Zhou},
  \citenamefont {Long}, \citenamefont {Tang}, \citenamefont {Chen},
  \citenamefont {Gao}, \citenamefont {Peng}, \citenamefont {Duan},
  \citenamefont {Zhong}, \citenamefont {Xiong}, \citenamefont {Wang},
  \citenamefont {Zhang},\ and\ \citenamefont {Zhan}}]{Zhou2015a}%
  \BibitemOpen
  \bibfield  {author} {\bibinfo {author} {\bibfnamefont {L.}~\bibnamefont
  {Zhou}}, \bibinfo {author} {\bibfnamefont {S.}~\bibnamefont {Long}}, \bibinfo
  {author} {\bibfnamefont {B.}~\bibnamefont {Tang}}, \bibinfo {author}
  {\bibfnamefont {X.}~\bibnamefont {Chen}}, \bibinfo {author} {\bibfnamefont
  {F.}~\bibnamefont {Gao}}, \bibinfo {author} {\bibfnamefont {W.}~\bibnamefont
  {Peng}}, \bibinfo {author} {\bibfnamefont {W.}~\bibnamefont {Duan}}, \bibinfo
  {author} {\bibfnamefont {J.}~\bibnamefont {Zhong}}, \bibinfo {author}
  {\bibfnamefont {Z.}~\bibnamefont {Xiong}}, \bibinfo {author} {\bibfnamefont
  {J.}~\bibnamefont {Wang}}, \bibinfo {author} {\bibfnamefont {Y.}~\bibnamefont
  {Zhang}}, \ and\ \bibinfo {author} {\bibfnamefont {M.}~\bibnamefont {Zhan}},\
  }\bibfield  {title} {\enquote {\bibinfo {title} {Test of equivalence
  principle at 10-8level by a dual-species double-diffraction raman atom
  interferometer},}\ }\href {\doibase 10.1103/physrevlett.115.013004}
  {\bibfield  {journal} {\bibinfo  {journal} {Physical Review Letters}\
  }\textbf {\bibinfo {volume} {115}} (\bibinfo {year} {2015}),\
  10.1103/physrevlett.115.013004}\BibitemShut {NoStop}%
\bibitem [{\citenamefont {Barrett}\ \emph {et~al.}(2016)\citenamefont
  {Barrett}, \citenamefont {Antoni-Micollier}, \citenamefont {Chichet},
  \citenamefont {Battelier}, \citenamefont {L{\'{e}}v{\`{e}}que}, \citenamefont
  {Landragin},\ and\ \citenamefont {Bouyer}}]{Barrett2016}%
  \BibitemOpen
  \bibfield  {author} {\bibinfo {author} {\bibfnamefont {Brynle}\ \bibnamefont
  {Barrett}}, \bibinfo {author} {\bibfnamefont {Laura}\ \bibnamefont
  {Antoni-Micollier}}, \bibinfo {author} {\bibfnamefont {Laure}\ \bibnamefont
  {Chichet}}, \bibinfo {author} {\bibfnamefont {Baptiste}\ \bibnamefont
  {Battelier}}, \bibinfo {author} {\bibfnamefont {Thomas}\ \bibnamefont
  {L{\'{e}}v{\`{e}}que}}, \bibinfo {author} {\bibfnamefont {Arnaud}\
  \bibnamefont {Landragin}}, \ and\ \bibinfo {author} {\bibfnamefont
  {Philippe}\ \bibnamefont {Bouyer}},\ }\bibfield  {title} {\enquote {\bibinfo
  {title} {Dual matter-wave inertial sensors in weightlessness},}\ }\href
  {\doibase 10.1038/ncomms13786} {\bibfield  {journal} {\bibinfo  {journal}
  {Nature Communications}\ }\textbf {\bibinfo {volume} {7}} (\bibinfo {year}
  {2016}),\ 10.1038/ncomms13786}\BibitemShut {NoStop}%
\bibitem [{\citenamefont {Parker}\ \emph {et~al.}(2018)\citenamefont {Parker},
  \citenamefont {Yu}, \citenamefont {Zhong}, \citenamefont {Estey},\ and\
  \citenamefont {M\"{u}ller}}]{Parker2018}%
  \BibitemOpen
  \bibfield  {author} {\bibinfo {author} {\bibfnamefont {Richard~H.}\
  \bibnamefont {Parker}}, \bibinfo {author} {\bibfnamefont {Chenghui}\
  \bibnamefont {Yu}}, \bibinfo {author} {\bibfnamefont {Weicheng}\ \bibnamefont
  {Zhong}}, \bibinfo {author} {\bibfnamefont {Brian}\ \bibnamefont {Estey}}, \
  and\ \bibinfo {author} {\bibfnamefont {Holger}\ \bibnamefont {M\"{u}ller}},\
  }\bibfield  {title} {\enquote {\bibinfo {title} {Measurement of the
  fine-structure constant as a test of the standard model},}\ }\href {\doibase
  10.1126/science.aap7706} {\bibfield  {journal} {\bibinfo  {journal}
  {Science}\ }\textbf {\bibinfo {volume} {360}} (\bibinfo {year} {2018}),\
  10.1126/science.aap7706}\BibitemShut {NoStop}%
\bibitem [{\citenamefont {Becker}\ \emph {et~al.}(2018)\citenamefont {Becker},
  \citenamefont {Lachmann}, \citenamefont {Seidel}, \citenamefont {Ahlers},
  \citenamefont {Dinkelaker}, \citenamefont {Grosse}, \citenamefont {Hellmig},
  \citenamefont {M{\"u}ntinga}, \citenamefont {Schkolnik}, \citenamefont
  {Wendrich}, \citenamefont {Wenzlawski}, \citenamefont {Weps}, \citenamefont
  {Corgier}, \citenamefont {Franz}, \citenamefont {Gaaloul}, \citenamefont
  {Herr}, \citenamefont {L{\"u}dtke}, \citenamefont {Popp}, \citenamefont
  {Amri}, \citenamefont {Duncker}, \citenamefont {Erbe}, \citenamefont
  {Kohfeldt}, \citenamefont {Kubelka-Lange}, \citenamefont {Braxmaier},
  \citenamefont {Charron}, \citenamefont {Ertmer}, \citenamefont {Krutzik},
  \citenamefont {L{\"a}mmerzahl}, \citenamefont {Peters}, \citenamefont
  {Schleich}, \citenamefont {Sengstock}, \citenamefont {Walser}, \citenamefont
  {Wicht}, \citenamefont {Windpassinger},\ and\ \citenamefont
  {Rasel}}]{Becker2018}%
  \BibitemOpen
  \bibfield  {author} {\bibinfo {author} {\bibfnamefont {Dennis}\ \bibnamefont
  {Becker}}, \bibinfo {author} {\bibfnamefont {Maike~D.}\ \bibnamefont
  {Lachmann}}, \bibinfo {author} {\bibfnamefont {Stephan~T.}\ \bibnamefont
  {Seidel}}, \bibinfo {author} {\bibfnamefont {Holger}\ \bibnamefont {Ahlers}},
  \bibinfo {author} {\bibfnamefont {Aline~N.}\ \bibnamefont {Dinkelaker}},
  \bibinfo {author} {\bibfnamefont {Jens}\ \bibnamefont {Grosse}}, \bibinfo
  {author} {\bibfnamefont {Ortwin}\ \bibnamefont {Hellmig}}, \bibinfo {author}
  {\bibfnamefont {Hauke}\ \bibnamefont {M{\"u}ntinga}}, \bibinfo {author}
  {\bibfnamefont {Vladimir}\ \bibnamefont {Schkolnik}}, \bibinfo {author}
  {\bibfnamefont {Thijs}\ \bibnamefont {Wendrich}}, \bibinfo {author}
  {\bibfnamefont {Andr{\'e}}\ \bibnamefont {Wenzlawski}}, \bibinfo {author}
  {\bibfnamefont {Benjamin}\ \bibnamefont {Weps}}, \bibinfo {author}
  {\bibfnamefont {Robin}\ \bibnamefont {Corgier}}, \bibinfo {author}
  {\bibfnamefont {Tobias}\ \bibnamefont {Franz}}, \bibinfo {author}
  {\bibfnamefont {Naceur}\ \bibnamefont {Gaaloul}}, \bibinfo {author}
  {\bibfnamefont {Waldemar}\ \bibnamefont {Herr}}, \bibinfo {author}
  {\bibfnamefont {Daniel}\ \bibnamefont {L{\"u}dtke}}, \bibinfo {author}
  {\bibfnamefont {Manuel}\ \bibnamefont {Popp}}, \bibinfo {author}
  {\bibfnamefont {Sirine}\ \bibnamefont {Amri}}, \bibinfo {author}
  {\bibfnamefont {Hannes}\ \bibnamefont {Duncker}}, \bibinfo {author}
  {\bibfnamefont {Maik}\ \bibnamefont {Erbe}}, \bibinfo {author} {\bibfnamefont
  {Anja}\ \bibnamefont {Kohfeldt}}, \bibinfo {author} {\bibfnamefont
  {Andr{\'e}}\ \bibnamefont {Kubelka-Lange}}, \bibinfo {author} {\bibfnamefont
  {Claus}\ \bibnamefont {Braxmaier}}, \bibinfo {author} {\bibfnamefont {Eric}\
  \bibnamefont {Charron}}, \bibinfo {author} {\bibfnamefont {Wolfgang}\
  \bibnamefont {Ertmer}}, \bibinfo {author} {\bibfnamefont {Markus}\
  \bibnamefont {Krutzik}}, \bibinfo {author} {\bibfnamefont {Claus}\
  \bibnamefont {L{\"a}mmerzahl}}, \bibinfo {author} {\bibfnamefont {Achim}\
  \bibnamefont {Peters}}, \bibinfo {author} {\bibfnamefont {Wolfgang~P.}\
  \bibnamefont {Schleich}}, \bibinfo {author} {\bibfnamefont {Klaus}\
  \bibnamefont {Sengstock}}, \bibinfo {author} {\bibfnamefont {Reinhold}\
  \bibnamefont {Walser}}, \bibinfo {author} {\bibfnamefont {Andreas}\
  \bibnamefont {Wicht}}, \bibinfo {author} {\bibfnamefont {Patrick}\
  \bibnamefont {Windpassinger}}, \ and\ \bibinfo {author} {\bibfnamefont
  {Ernst~M.}\ \bibnamefont {Rasel}},\ }\bibfield  {title} {\enquote {\bibinfo
  {title} {Space-borne bose-einstein condensation for precision
  interferometry},}\ }\href {\doibase 10.1038/s41586-018-0605-1} {\bibfield
  {journal} {\bibinfo  {journal} {Nature}\ }\textbf {\bibinfo {volume} {562}}
  (\bibinfo {year} {2018}),\ 10.1038/s41586-018-0605-1}\BibitemShut {NoStop}%
\bibitem [{\citenamefont {Bongs}\ \emph {et~al.}(2019)\citenamefont {Bongs},
  \citenamefont {Holynski}, \citenamefont {Vovrosh}, \citenamefont {Bouyer},
  \citenamefont {Condon}, \citenamefont {Rasel}, \citenamefont {Schubert},
  \citenamefont {Schleich},\ and\ \citenamefont {Roura}}]{Bongs2019}%
  \BibitemOpen
  \bibfield  {author} {\bibinfo {author} {\bibfnamefont {Kai}\ \bibnamefont
  {Bongs}}, \bibinfo {author} {\bibfnamefont {Michael}\ \bibnamefont
  {Holynski}}, \bibinfo {author} {\bibfnamefont {Jamie}\ \bibnamefont
  {Vovrosh}}, \bibinfo {author} {\bibfnamefont {Philippe}\ \bibnamefont
  {Bouyer}}, \bibinfo {author} {\bibfnamefont {Gabriel}\ \bibnamefont
  {Condon}}, \bibinfo {author} {\bibfnamefont {Ernst}\ \bibnamefont {Rasel}},
  \bibinfo {author} {\bibfnamefont {Christian}\ \bibnamefont {Schubert}},
  \bibinfo {author} {\bibfnamefont {Wolfgang~P.}\ \bibnamefont {Schleich}}, \
  and\ \bibinfo {author} {\bibfnamefont {Albert}\ \bibnamefont {Roura}},\
  }\bibfield  {title} {\enquote {\bibinfo {title} {Taking atom interferometric
  quantum sensors from the laboratory to real-world applications},}\ }\href
  {\doibase 10.1038/s42254-019-0117-4} {\bibfield  {journal} {\bibinfo
  {journal} {Nature Reviews Physics}\ }\textbf {\bibinfo {volume} {1}}
  (\bibinfo {year} {2019}),\ 10.1038/s42254-019-0117-4}\BibitemShut {NoStop}%
\bibitem [{\citenamefont {Wu}(2009)}]{Wu2009}%
  \BibitemOpen
  \bibfield  {author} {\bibinfo {author} {\bibfnamefont {Xinan}\ \bibnamefont
  {Wu}},\ }\emph {\bibinfo {title} {Gravity Gradient Survey with a Mobile Atom
  Interferometer}},\ \href@noop {} {Ph.D. thesis},\ \bibinfo  {school}
  {Stanford} (\bibinfo {year} {2009})\BibitemShut {NoStop}%
\bibitem [{\citenamefont {Farah}\ \emph {et~al.}(2014)\citenamefont {Farah},
  \citenamefont {Guerlin}, \citenamefont {Landragin}, \citenamefont {Bouyer},
  \citenamefont {Gaffet}, \citenamefont {Santos},\ and\ \citenamefont
  {Merlet}}]{Farah2014}%
  \BibitemOpen
  \bibfield  {author} {\bibinfo {author} {\bibfnamefont {T.}~\bibnamefont
  {Farah}}, \bibinfo {author} {\bibfnamefont {C.}~\bibnamefont {Guerlin}},
  \bibinfo {author} {\bibfnamefont {A.}~\bibnamefont {Landragin}}, \bibinfo
  {author} {\bibfnamefont {Ph.}\ \bibnamefont {Bouyer}}, \bibinfo {author}
  {\bibfnamefont {S.}~\bibnamefont {Gaffet}}, \bibinfo {author} {\bibfnamefont
  {F.~Pereira~Dos}\ \bibnamefont {Santos}}, \ and\ \bibinfo {author}
  {\bibfnamefont {S.}~\bibnamefont {Merlet}},\ }\bibfield  {title} {\enquote
  {\bibinfo {title} {Underground operation at best sensitivity of the mobile
  {LNE}-{SYRTE} cold atom gravimeter},}\ }\href {\doibase
  10.1134/s2075108714040051} {\bibfield  {journal} {\bibinfo  {journal}
  {Gyroscopy and Navigation}\ }\textbf {\bibinfo {volume} {5}} (\bibinfo {year}
  {2014}),\ 10.1134/s2075108714040051}\BibitemShut {NoStop}%
\bibitem [{\citenamefont {Freier}\ \emph {et~al.}(2016)\citenamefont {Freier},
  \citenamefont {Hauth}, \citenamefont {Schkolnik}, \citenamefont {Leykauf},
  \citenamefont {Schilling}, \citenamefont {Wziontek}, \citenamefont
  {Scherneck}, \citenamefont {M\"{u}ller},\ and\ \citenamefont
  {Peters}}]{Freier2016}%
  \BibitemOpen
  \bibfield  {author} {\bibinfo {author} {\bibfnamefont {C}~\bibnamefont
  {Freier}}, \bibinfo {author} {\bibfnamefont {M}~\bibnamefont {Hauth}},
  \bibinfo {author} {\bibfnamefont {V}~\bibnamefont {Schkolnik}}, \bibinfo
  {author} {\bibfnamefont {B}~\bibnamefont {Leykauf}}, \bibinfo {author}
  {\bibfnamefont {M}~\bibnamefont {Schilling}}, \bibinfo {author}
  {\bibfnamefont {H}~\bibnamefont {Wziontek}}, \bibinfo {author} {\bibfnamefont
  {H-G}\ \bibnamefont {Scherneck}}, \bibinfo {author} {\bibfnamefont
  {J}~\bibnamefont {M\"{u}ller}}, \ and\ \bibinfo {author} {\bibfnamefont
  {A}~\bibnamefont {Peters}},\ }\bibfield  {title} {\enquote {\bibinfo {title}
  {Mobile quantum gravity sensor with unprecedented stability},}\ }\href
  {\doibase 10.1088/1742-6596/723/1/012050} {\bibfield  {journal} {\bibinfo
  {journal} {Journal of Physics: Conference Series}\ }\textbf {\bibinfo
  {volume} {723}} (\bibinfo {year} {2016}),\
  10.1088/1742-6596/723/1/012050}\BibitemShut {NoStop}%
\bibitem [{\citenamefont {M\'{e}noret}\ \emph {et~al.}(2018)\citenamefont
  {M\'{e}noret}, \citenamefont {Vermeulen}, \citenamefont {Le~Moigne},
  \citenamefont {Bonvalot}, \citenamefont {Bouyer}, \citenamefont {Landragin},\
  and\ \citenamefont {Desruelle}}]{Menoret2018}%
  \BibitemOpen
  \bibfield  {author} {\bibinfo {author} {\bibfnamefont {Vincent}\ \bibnamefont
  {M\'{e}noret}}, \bibinfo {author} {\bibfnamefont {Pierre}\ \bibnamefont
  {Vermeulen}}, \bibinfo {author} {\bibfnamefont {Nicolas}\ \bibnamefont
  {Le~Moigne}}, \bibinfo {author} {\bibfnamefont {Sylvain}\ \bibnamefont
  {Bonvalot}}, \bibinfo {author} {\bibfnamefont {Philippe}\ \bibnamefont
  {Bouyer}}, \bibinfo {author} {\bibfnamefont {Arnaud}\ \bibnamefont
  {Landragin}}, \ and\ \bibinfo {author} {\bibfnamefont {Bruno}\ \bibnamefont
  {Desruelle}},\ }\bibfield  {title} {\enquote {\bibinfo {title} {Gravity
  measurements below 10-9 g with a transportable absolute quantum
  gravimeter},}\ }\href {https://doi.org/10.1038/s41598-018-30608-1} {\bibfield
   {journal} {\bibinfo  {journal} {Scientific Reports}\ }\textbf {\bibinfo
  {volume} {8}} (\bibinfo {year} {2018})}\BibitemShut {NoStop}%
\bibitem [{\citenamefont {Bidel}\ \emph {et~al.}(2018)\citenamefont {Bidel},
  \citenamefont {Zahzam}, \citenamefont {Blanchard}, \citenamefont {Bonnin},
  \citenamefont {Cadoret}, \citenamefont {Bresson}, \citenamefont {Rouxel},\
  and\ \citenamefont {Lequentrec-Lalancette}}]{Bidel2018}%
  \BibitemOpen
  \bibfield  {author} {\bibinfo {author} {\bibfnamefont {Y.}~\bibnamefont
  {Bidel}}, \bibinfo {author} {\bibfnamefont {N.}~\bibnamefont {Zahzam}},
  \bibinfo {author} {\bibfnamefont {C.}~\bibnamefont {Blanchard}}, \bibinfo
  {author} {\bibfnamefont {A.}~\bibnamefont {Bonnin}}, \bibinfo {author}
  {\bibfnamefont {M.}~\bibnamefont {Cadoret}}, \bibinfo {author} {\bibfnamefont
  {A.}~\bibnamefont {Bresson}}, \bibinfo {author} {\bibfnamefont
  {D.}~\bibnamefont {Rouxel}}, \ and\ \bibinfo {author} {\bibfnamefont {M.~F.}\
  \bibnamefont {Lequentrec-Lalancette}},\ }\bibfield  {title} {\enquote
  {\bibinfo {title} {Absolute marine gravimetry with matter-wave
  interferometry},}\ }\href {\doibase 10.1038/s41467-018-03040-2} {\bibfield
  {journal} {\bibinfo  {journal} {Nature Communications}\ }\textbf {\bibinfo
  {volume} {9}} (\bibinfo {year} {2018}),\
  10.1038/s41467-018-03040-2}\BibitemShut {NoStop}%
\bibitem [{\citenamefont {Wu}\ \emph {et~al.}(2019)\citenamefont {Wu},
  \citenamefont {Pagel}, \citenamefont {Malek}, \citenamefont {Nguyen},
  \citenamefont {Zi}, \citenamefont {Scheirer},\ and\ \citenamefont
  {M{\"u}ller}}]{Wu2019}%
  \BibitemOpen
  \bibfield  {author} {\bibinfo {author} {\bibfnamefont {Xuejian}\ \bibnamefont
  {Wu}}, \bibinfo {author} {\bibfnamefont {Zachary}\ \bibnamefont {Pagel}},
  \bibinfo {author} {\bibfnamefont {Bola~S.}\ \bibnamefont {Malek}}, \bibinfo
  {author} {\bibfnamefont {Timothy~H.}\ \bibnamefont {Nguyen}}, \bibinfo
  {author} {\bibfnamefont {Fei}\ \bibnamefont {Zi}}, \bibinfo {author}
  {\bibfnamefont {Daniel~S.}\ \bibnamefont {Scheirer}}, \ and\ \bibinfo
  {author} {\bibfnamefont {Holger}\ \bibnamefont {M{\"u}ller}},\ }\bibfield
  {title} {\enquote {\bibinfo {title} {Gravity surveys using a mobile atom
  interferometer},}\ }\href {\doibase 10.1126/sciadv.aax0800} {\bibfield
  {journal} {\bibinfo  {journal} {Science Advances}\ }\textbf {\bibinfo
  {volume} {5}} (\bibinfo {year} {2019}),\ 10.1126/sciadv.aax0800}\BibitemShut
  {NoStop}%
\bibitem [{\citenamefont {Bidel}\ \emph {et~al.}()\citenamefont {Bidel},
  \citenamefont {Zahzam}, \citenamefont {Bresson}, \citenamefont {Blanchard},
  \citenamefont {Cadoret}, \citenamefont {Olesen},\ and\ \citenamefont
  {Forsberg}}]{Bidel2019}%
  \BibitemOpen
  \bibfield  {author} {\bibinfo {author} {\bibfnamefont {Yannick}\ \bibnamefont
  {Bidel}}, \bibinfo {author} {\bibfnamefont {Nassim}\ \bibnamefont {Zahzam}},
  \bibinfo {author} {\bibfnamefont {Alexandre}\ \bibnamefont {Bresson}},
  \bibinfo {author} {\bibfnamefont {C{\'e}dric}\ \bibnamefont {Blanchard}},
  \bibinfo {author} {\bibfnamefont {Malo}\ \bibnamefont {Cadoret}}, \bibinfo
  {author} {\bibfnamefont {Arne~V.}\ \bibnamefont {Olesen}}, \ and\ \bibinfo
  {author} {\bibfnamefont {Ren{\'e}}\ \bibnamefont {Forsberg}},\ }\bibfield
  {title} {\enquote {\bibinfo {title} {Absolute airborne gravimetry with a cold
  atom sensor},}\ }\href@noop {} {\ }\Eprint
  {http://arxiv.org/abs/http://arxiv.org/abs/1910.06666v1}
  {http://arxiv.org/abs/1910.06666v1} \BibitemShut {NoStop}%
\bibitem [{\citenamefont {Barrett}\ \emph {et~al.}(2014)\citenamefont
  {Barrett}, \citenamefont {Geiger}, \citenamefont {Dutta}, \citenamefont
  {Meunier}, \citenamefont {Canuel}, \citenamefont {Gauguet}, \citenamefont
  {Bouyer},\ and\ \citenamefont {Landragin}}]{Barrett2014}%
  \BibitemOpen
  \bibfield  {author} {\bibinfo {author} {\bibfnamefont {Brynle}\ \bibnamefont
  {Barrett}}, \bibinfo {author} {\bibfnamefont {R{\'{e}}my}\ \bibnamefont
  {Geiger}}, \bibinfo {author} {\bibfnamefont {Indranil}\ \bibnamefont
  {Dutta}}, \bibinfo {author} {\bibfnamefont {Matthieu}\ \bibnamefont
  {Meunier}}, \bibinfo {author} {\bibfnamefont {Benjamin}\ \bibnamefont
  {Canuel}}, \bibinfo {author} {\bibfnamefont {Alexandre}\ \bibnamefont
  {Gauguet}}, \bibinfo {author} {\bibfnamefont {Philippe}\ \bibnamefont
  {Bouyer}}, \ and\ \bibinfo {author} {\bibfnamefont {Arnaud}\ \bibnamefont
  {Landragin}},\ }\bibfield  {title} {\enquote {\bibinfo {title} {The sagnac
  effect: 20 years of development in matter-wave interferometry},}\ }\href
  {\doibase 10.1016/j.crhy.2014.10.009} {\bibfield  {journal} {\bibinfo
  {journal} {Comptes Rendus Physique}\ }\textbf {\bibinfo {volume} {15}}
  (\bibinfo {year} {2014}),\ 10.1016/j.crhy.2014.10.009}\BibitemShut {NoStop}%
\bibitem [{\citenamefont {Gustavson}\ \emph {et~al.}(1997)\citenamefont
  {Gustavson}, \citenamefont {Bouyer},\ and\ \citenamefont
  {Kasevich}}]{Gustavson1997}%
  \BibitemOpen
  \bibfield  {author} {\bibinfo {author} {\bibfnamefont {T.~L.}\ \bibnamefont
  {Gustavson}}, \bibinfo {author} {\bibfnamefont {P.}~\bibnamefont {Bouyer}}, \
  and\ \bibinfo {author} {\bibfnamefont {M.~A.}\ \bibnamefont {Kasevich}},\
  }\bibfield  {title} {\enquote {\bibinfo {title} {Precision rotation
  measurements with an atom interferometer gyroscope},}\ }\href {\doibase
  10.1103/physrevlett.78.2046} {\bibfield  {journal} {\bibinfo  {journal}
  {Physical Review Letters}\ }\textbf {\bibinfo {volume} {78}} (\bibinfo {year}
  {1997}),\ 10.1103/physrevlett.78.2046}\BibitemShut {NoStop}%
\bibitem [{\citenamefont {Stockton}\ \emph {et~al.}(2011)\citenamefont
  {Stockton}, \citenamefont {Takase},\ and\ \citenamefont
  {Kasevich}}]{Stockton2011}%
  \BibitemOpen
  \bibfield  {author} {\bibinfo {author} {\bibfnamefont {J.~K.}\ \bibnamefont
  {Stockton}}, \bibinfo {author} {\bibfnamefont {K.}~\bibnamefont {Takase}}, \
  and\ \bibinfo {author} {\bibfnamefont {M.~A.}\ \bibnamefont {Kasevich}},\
  }\bibfield  {title} {\enquote {\bibinfo {title} {Absolute geodetic rotation
  measurement using atom interferometry},}\ }\href {\doibase
  10.1103/physrevlett.107.133001} {\bibfield  {journal} {\bibinfo  {journal}
  {Physical Review Letters}\ }\textbf {\bibinfo {volume} {107}} (\bibinfo
  {year} {2011}),\ 10.1103/physrevlett.107.133001}\BibitemShut {NoStop}%
\bibitem [{\citenamefont {Dickerson}\ \emph {et~al.}(2013)\citenamefont
  {Dickerson}, \citenamefont {Hogan}, \citenamefont {Sugarbaker}, \citenamefont
  {Johnson},\ and\ \citenamefont {Kasevich}}]{Dickerson2013}%
  \BibitemOpen
  \bibfield  {author} {\bibinfo {author} {\bibfnamefont {Susannah~M.}\
  \bibnamefont {Dickerson}}, \bibinfo {author} {\bibfnamefont {Jason~M.}\
  \bibnamefont {Hogan}}, \bibinfo {author} {\bibfnamefont {Alex}\ \bibnamefont
  {Sugarbaker}}, \bibinfo {author} {\bibfnamefont {David M.~S.}\ \bibnamefont
  {Johnson}}, \ and\ \bibinfo {author} {\bibfnamefont {Mark~A.}\ \bibnamefont
  {Kasevich}},\ }\bibfield  {title} {\enquote {\bibinfo {title} {Multiaxis
  inertial sensing with long-time point source atom interferometry},}\ }\href
  {\doibase 10.1103/physrevlett.111.083001} {\bibfield  {journal} {\bibinfo
  {journal} {Physical Review Letters}\ }\textbf {\bibinfo {volume} {111}}
  (\bibinfo {year} {2013}),\ 10.1103/physrevlett.111.083001}\BibitemShut
  {NoStop}%
\bibitem [{\citenamefont {Savoie}\ \emph {et~al.}(2018)\citenamefont {Savoie},
  \citenamefont {Altorio}, \citenamefont {Fang}, \citenamefont {Sidorenkov},
  \citenamefont {Geiger},\ and\ \citenamefont {Landragin}}]{Savoie2018}%
  \BibitemOpen
  \bibfield  {author} {\bibinfo {author} {\bibfnamefont {D.}~\bibnamefont
  {Savoie}}, \bibinfo {author} {\bibfnamefont {M.}~\bibnamefont {Altorio}},
  \bibinfo {author} {\bibfnamefont {B.}~\bibnamefont {Fang}}, \bibinfo {author}
  {\bibfnamefont {L.~A.}\ \bibnamefont {Sidorenkov}}, \bibinfo {author}
  {\bibfnamefont {R.}~\bibnamefont {Geiger}}, \ and\ \bibinfo {author}
  {\bibfnamefont {A.}~\bibnamefont {Landragin}},\ }\bibfield  {title} {\enquote
  {\bibinfo {title} {Interleaved atom interferometry for high-sensitivity
  inertial measurements},}\ }\href {\doibase 10.1126/sciadv.aau7948} {\bibfield
   {journal} {\bibinfo  {journal} {Science Advances}\ }\textbf {\bibinfo
  {volume} {4}} (\bibinfo {year} {2018}),\ 10.1126/sciadv.aau7948}\BibitemShut
  {NoStop}%
\bibitem [{\citenamefont {Chen}\ \emph {et~al.}(2019)\citenamefont {Chen},
  \citenamefont {Hansen}, \citenamefont {Hoth}, \citenamefont {Ivanov},
  \citenamefont {Pelle}, \citenamefont {Kitching},\ and\ \citenamefont
  {Donley}}]{Chen2019}%
  \BibitemOpen
  \bibfield  {author} {\bibinfo {author} {\bibfnamefont {Yun-Jhih}\
  \bibnamefont {Chen}}, \bibinfo {author} {\bibfnamefont {Azure}\ \bibnamefont
  {Hansen}}, \bibinfo {author} {\bibfnamefont {Gregory~W.}\ \bibnamefont
  {Hoth}}, \bibinfo {author} {\bibfnamefont {Eugene}\ \bibnamefont {Ivanov}},
  \bibinfo {author} {\bibfnamefont {Bruno}\ \bibnamefont {Pelle}}, \bibinfo
  {author} {\bibfnamefont {John}\ \bibnamefont {Kitching}}, \ and\ \bibinfo
  {author} {\bibfnamefont {Elizabeth~A.}\ \bibnamefont {Donley}},\ }\bibfield
  {title} {\enquote {\bibinfo {title} {Single-source multiaxis cold-atom
  interferometer in a centimeter-scale cell},}\ }\href {\doibase
  10.1103/physrevapplied.12.014019} {\bibfield  {journal} {\bibinfo  {journal}
  {Physical Review Applied}\ }\textbf {\bibinfo {volume} {12}} (\bibinfo {year}
  {2019}),\ 10.1103/physrevapplied.12.014019}\BibitemShut {NoStop}%
\bibitem [{\citenamefont {Gauguet}\ \emph {et~al.}(2009)\citenamefont
  {Gauguet}, \citenamefont {Canuel}, \citenamefont {L{\'{e}}v{\`{e}}que},
  \citenamefont {Chaibi},\ and\ \citenamefont {Landragin}}]{Gauguet2009}%
  \BibitemOpen
  \bibfield  {author} {\bibinfo {author} {\bibfnamefont {A.}~\bibnamefont
  {Gauguet}}, \bibinfo {author} {\bibfnamefont {B.}~\bibnamefont {Canuel}},
  \bibinfo {author} {\bibfnamefont {T.}~\bibnamefont {L{\'{e}}v{\`{e}}que}},
  \bibinfo {author} {\bibfnamefont {W.}~\bibnamefont {Chaibi}}, \ and\ \bibinfo
  {author} {\bibfnamefont {A.}~\bibnamefont {Landragin}},\ }\bibfield  {title}
  {\enquote {\bibinfo {title} {Characterization and limits of a cold-atom
  sagnac interferometer},}\ }\href {\doibase 10.1103/physreva.80.063604}
  {\bibfield  {journal} {\bibinfo  {journal} {Physical Review A}\ }\textbf
  {\bibinfo {volume} {80}} (\bibinfo {year} {2009}),\
  10.1103/physreva.80.063604}\BibitemShut {NoStop}%
\bibitem [{\citenamefont {Sugarbaker}\ \emph {et~al.}(2013)\citenamefont
  {Sugarbaker}, \citenamefont {Dickerson}, \citenamefont {Hogan}, \citenamefont
  {Johnson},\ and\ \citenamefont {Kasevich}}]{Sugarbaker2013}%
  \BibitemOpen
  \bibfield  {author} {\bibinfo {author} {\bibfnamefont {Alex}\ \bibnamefont
  {Sugarbaker}}, \bibinfo {author} {\bibfnamefont {Susannah~M.}\ \bibnamefont
  {Dickerson}}, \bibinfo {author} {\bibfnamefont {Jason~M.}\ \bibnamefont
  {Hogan}}, \bibinfo {author} {\bibfnamefont {David M.~S.}\ \bibnamefont
  {Johnson}}, \ and\ \bibinfo {author} {\bibfnamefont {Mark~A.}\ \bibnamefont
  {Kasevich}},\ }\bibfield  {title} {\enquote {\bibinfo {title} {Enhanced atom
  interferometer readout through the application of phase shear},}\ }\href
  {\doibase 10.1103/physrevlett.111.113002} {\bibfield  {journal} {\bibinfo
  {journal} {Physical Review Letters}\ }\textbf {\bibinfo {volume} {111}}
  (\bibinfo {year} {2013}),\ 10.1103/physrevlett.111.113002}\BibitemShut
  {NoStop}%
\bibitem [{\citenamefont {Jekeli}(2005)}]{JEKELI2005}%
  \BibitemOpen
  \bibfield  {author} {\bibinfo {author} {\bibfnamefont {C.}~\bibnamefont
  {Jekeli}},\ }\bibfield  {title} {\enquote {\bibinfo {title} {Navigation error
  analysis of atom interferometer inertial sensor},}\ }\href {\doibase
  10.1002/j.2161-4296.2005.tb01726.x} {\bibfield  {journal} {\bibinfo
  {journal} {Navigation}\ }\textbf {\bibinfo {volume} {52}},\ \bibinfo {pages}
  {1--14} (\bibinfo {year} {2005})}\BibitemShut {NoStop}%
\bibitem [{\citenamefont {Canuel}\ \emph {et~al.}(2006)\citenamefont {Canuel},
  \citenamefont {Leduc}, \citenamefont {Holleville}, \citenamefont {Gauguet},
  \citenamefont {Fils}, \citenamefont {Virdis}, \citenamefont {Clairon},
  \citenamefont {Dimarcq}, \citenamefont {Bord{\'{e}}}, \citenamefont
  {Landragin},\ and\ \citenamefont {Bouyer}}]{Canuel2006}%
  \BibitemOpen
  \bibfield  {author} {\bibinfo {author} {\bibfnamefont {B.}~\bibnamefont
  {Canuel}}, \bibinfo {author} {\bibfnamefont {F.}~\bibnamefont {Leduc}},
  \bibinfo {author} {\bibfnamefont {D.}~\bibnamefont {Holleville}}, \bibinfo
  {author} {\bibfnamefont {A.}~\bibnamefont {Gauguet}}, \bibinfo {author}
  {\bibfnamefont {J.}~\bibnamefont {Fils}}, \bibinfo {author} {\bibfnamefont
  {A.}~\bibnamefont {Virdis}}, \bibinfo {author} {\bibfnamefont
  {A.}~\bibnamefont {Clairon}}, \bibinfo {author} {\bibfnamefont
  {N.}~\bibnamefont {Dimarcq}}, \bibinfo {author} {\bibfnamefont {Ch.~J.}\
  \bibnamefont {Bord{\'{e}}}}, \bibinfo {author} {\bibfnamefont
  {A.}~\bibnamefont {Landragin}}, \ and\ \bibinfo {author} {\bibfnamefont
  {P.}~\bibnamefont {Bouyer}},\ }\bibfield  {title} {\enquote {\bibinfo {title}
  {Six-axis inertial sensor using cold-atom interferometry},}\ }\href {\doibase
  10.1103/physrevlett.97.010402} {\bibfield  {journal} {\bibinfo  {journal}
  {Physical Review Letters}\ }\textbf {\bibinfo {volume} {97}} (\bibinfo {year}
  {2006}),\ 10.1103/physrevlett.97.010402}\BibitemShut {NoStop}%
\bibitem [{\citenamefont {Rakholia}\ \emph {et~al.}(2014)\citenamefont
  {Rakholia}, \citenamefont {McGuinness},\ and\ \citenamefont
  {Biedermann}}]{Rakholia2014}%
  \BibitemOpen
  \bibfield  {author} {\bibinfo {author} {\bibfnamefont {Akash~V.}\
  \bibnamefont {Rakholia}}, \bibinfo {author} {\bibfnamefont {Hayden~J.}\
  \bibnamefont {McGuinness}}, \ and\ \bibinfo {author} {\bibfnamefont
  {Grant~W.}\ \bibnamefont {Biedermann}},\ }\bibfield  {title} {\enquote
  {\bibinfo {title} {Dual-axis high-data-rate atom interferometer via cold
  ensemble exchange},}\ }\href {\doibase 10.1103/physrevapplied.2.054012}
  {\bibfield  {journal} {\bibinfo  {journal} {Physical Review Applied}\
  }\textbf {\bibinfo {volume} {2}} (\bibinfo {year} {2014}),\
  10.1103/physrevapplied.2.054012}\BibitemShut {NoStop}%
\bibitem [{\citenamefont {Durfee}\ \emph {et~al.}(2006)\citenamefont {Durfee},
  \citenamefont {Shaham},\ and\ \citenamefont {Kasevich}}]{Durfee2006}%
  \BibitemOpen
  \bibfield  {author} {\bibinfo {author} {\bibfnamefont {D.~S.}\ \bibnamefont
  {Durfee}}, \bibinfo {author} {\bibfnamefont {Y.~K.}\ \bibnamefont {Shaham}},
  \ and\ \bibinfo {author} {\bibfnamefont {M.~A.}\ \bibnamefont {Kasevich}},\
  }\bibfield  {title} {\enquote {\bibinfo {title} {Long-term stability of an
  area-reversible atom-interferometer sagnac gyroscope},}\ }\href {\doibase
  10.1103/physrevlett.97.240801} {\bibfield  {journal} {\bibinfo  {journal}
  {Physical Review Letters}\ }\textbf {\bibinfo {volume} {97}} (\bibinfo {year}
  {2006}),\ 10.1103/physrevlett.97.240801}\BibitemShut {NoStop}%
\bibitem [{\citenamefont {Lefevre}(2014)}]{Lefevre2014}%
  \BibitemOpen
  \bibfield  {author} {\bibinfo {author} {\bibfnamefont {H.~C.}\ \bibnamefont
  {Lefevre}},\ }\href@noop {} {\emph {\bibinfo {title} {The Fiber-Optic
  Gyroscope}}}\ (\bibinfo  {publisher} {Artech House, London, UK},\ \bibinfo
  {year} {2014})\BibitemShut {NoStop}%
\bibitem [{\citenamefont {Battelier}\ \emph {et~al.}(2016)\citenamefont
  {Battelier}, \citenamefont {Barrett}, \citenamefont {Fouch{\'{e}}},
  \citenamefont {Chichet}, \citenamefont {Antoni-Micollier}, \citenamefont
  {Porte}, \citenamefont {Napolitano}, \citenamefont {Lautier}, \citenamefont
  {Landragin},\ and\ \citenamefont {Bouyer}}]{Battelier2016}%
  \BibitemOpen
  \bibfield  {author} {\bibinfo {author} {\bibfnamefont {B.}~\bibnamefont
  {Battelier}}, \bibinfo {author} {\bibfnamefont {B.}~\bibnamefont {Barrett}},
  \bibinfo {author} {\bibfnamefont {L.}~\bibnamefont {Fouch{\'{e}}}}, \bibinfo
  {author} {\bibfnamefont {L.}~\bibnamefont {Chichet}}, \bibinfo {author}
  {\bibfnamefont {L.}~\bibnamefont {Antoni-Micollier}}, \bibinfo {author}
  {\bibfnamefont {H.}~\bibnamefont {Porte}}, \bibinfo {author} {\bibfnamefont
  {F.}~\bibnamefont {Napolitano}}, \bibinfo {author} {\bibfnamefont
  {J.}~\bibnamefont {Lautier}}, \bibinfo {author} {\bibfnamefont
  {A.}~\bibnamefont {Landragin}}, \ and\ \bibinfo {author} {\bibfnamefont
  {P.}~\bibnamefont {Bouyer}},\ }\bibfield  {title} {\enquote {\bibinfo {title}
  {Development of compact cold-atom sensors for inertial navigation},}\ }in\
  \href {\doibase 10.1117/12.2228351} {\emph {\bibinfo {booktitle} {Quantum
  Optics}}},\ \bibinfo {editor} {edited by\ \bibinfo {editor} {\bibfnamefont
  {Jürgen}\ \bibnamefont {Stuhler}}\ and\ \bibinfo {editor} {\bibfnamefont
  {Andrew~J.}\ \bibnamefont {Shields}}}\ (\bibinfo  {publisher} {{SPIE}},\
  \bibinfo {year} {2016})\BibitemShut {NoStop}%
\bibitem [{\citenamefont {Hoth}\ \emph {et~al.}(2016)\citenamefont {Hoth},
  \citenamefont {Pelle}, \citenamefont {Riedl}, \citenamefont {Kitching},\ and\
  \citenamefont {Donley}}]{Hoth2016}%
  \BibitemOpen
  \bibfield  {author} {\bibinfo {author} {\bibfnamefont {Gregory~W.}\
  \bibnamefont {Hoth}}, \bibinfo {author} {\bibfnamefont {Bruno}\ \bibnamefont
  {Pelle}}, \bibinfo {author} {\bibfnamefont {Stefan}\ \bibnamefont {Riedl}},
  \bibinfo {author} {\bibfnamefont {John}\ \bibnamefont {Kitching}}, \ and\
  \bibinfo {author} {\bibfnamefont {Elizabeth~A.}\ \bibnamefont {Donley}},\
  }\bibfield  {title} {\enquote {\bibinfo {title} {Point source atom
  interferometry with a cloud of finite size},}\ }\href {\doibase
  10.1063/1.4961527} {\bibfield  {journal} {\bibinfo  {journal} {Applied
  Physics Letters}\ }\textbf {\bibinfo {volume} {109}} (\bibinfo {year}
  {2016}),\ 10.1063/1.4961527}\BibitemShut {NoStop}%
\bibitem [{\citenamefont {Hoth}\ \emph {et~al.}(2017)\citenamefont {Hoth},
  \citenamefont {Pelle}, \citenamefont {Kitching},\ and\ \citenamefont
  {Donley}}]{Hoth2017}%
  \BibitemOpen
  \bibfield  {author} {\bibinfo {author} {\bibfnamefont {Gregory~W.}\
  \bibnamefont {Hoth}}, \bibinfo {author} {\bibfnamefont {Bruno}\ \bibnamefont
  {Pelle}}, \bibinfo {author} {\bibfnamefont {John}\ \bibnamefont {Kitching}},
  \ and\ \bibinfo {author} {\bibfnamefont {Elizabeth~A.}\ \bibnamefont
  {Donley}},\ }\bibfield  {title} {\enquote {\bibinfo {title} {Analytical tools
  for point source interferometry},}\ }in\ \href {\doibase 10.1117/12.2247688}
  {\emph {\bibinfo {booktitle} {Slow Light, Fast Light, and Opto-Atomic
  Precision Metrology X}}},\ \bibinfo {editor} {edited by\ \bibinfo {editor}
  {\bibfnamefont {Selim~M.}\ \bibnamefont {Shahriar}}\ and\ \bibinfo {editor}
  {\bibfnamefont {Jacob}\ \bibnamefont {Scheuer}}}\ (\bibinfo  {publisher}
  {{SPIE}},\ \bibinfo {year} {2017})\BibitemShut {NoStop}%
\bibitem [{\citenamefont {Kasevich}\ and\ \citenamefont
  {Chu}(1991)}]{Kasevich_1991}%
  \BibitemOpen
  \bibfield  {author} {\bibinfo {author} {\bibfnamefont {Mark}\ \bibnamefont
  {Kasevich}}\ and\ \bibinfo {author} {\bibfnamefont {Steven}\ \bibnamefont
  {Chu}},\ }\bibfield  {title} {\enquote {\bibinfo {title} {Atomic
  interferometry using stimulated raman transitions},}\ }\href {\doibase
  10.1103/physrevlett.67.181} {\bibfield  {journal} {\bibinfo  {journal}
  {Physical Review Letters}\ }\textbf {\bibinfo {volume} {67}} (\bibinfo {year}
  {1991}),\ 10.1103/physrevlett.67.181}\BibitemShut {NoStop}%
\bibitem [{\citenamefont {Kasevich}\ \emph {et~al.}(1991)\citenamefont
  {Kasevich}, \citenamefont {Weiss}, \citenamefont {Riis}, \citenamefont
  {Moler}, \citenamefont {Kasapi},\ and\ \citenamefont {Chu}}]{Kasevich1991a}%
  \BibitemOpen
  \bibfield  {author} {\bibinfo {author} {\bibfnamefont {Mark}\ \bibnamefont
  {Kasevich}}, \bibinfo {author} {\bibfnamefont {David~S.}\ \bibnamefont
  {Weiss}}, \bibinfo {author} {\bibfnamefont {Erling}\ \bibnamefont {Riis}},
  \bibinfo {author} {\bibfnamefont {Kathryn}\ \bibnamefont {Moler}}, \bibinfo
  {author} {\bibfnamefont {Steven}\ \bibnamefont {Kasapi}}, \ and\ \bibinfo
  {author} {\bibfnamefont {Steven}\ \bibnamefont {Chu}},\ }\bibfield  {title}
  {\enquote {\bibinfo {title} {Atomic velocity selection using stimulated raman
  transitions},}\ }\href {\doibase 10.1103/physrevlett.66.2297} {\bibfield
  {journal} {\bibinfo  {journal} {Physical Review Letters}\ }\textbf {\bibinfo
  {volume} {66}} (\bibinfo {year} {1991}),\
  10.1103/physrevlett.66.2297}\BibitemShut {NoStop}%
\bibitem [{\citenamefont {Yankelev}\ \emph {et~al.}(2019)\citenamefont
  {Yankelev}, \citenamefont {Avinadav}, \citenamefont {Davidson},\ and\
  \citenamefont {Firstenberg}}]{Yankelev2019}%
  \BibitemOpen
  \bibfield  {author} {\bibinfo {author} {\bibfnamefont {Dimitry}\ \bibnamefont
  {Yankelev}}, \bibinfo {author} {\bibfnamefont {Chen}\ \bibnamefont
  {Avinadav}}, \bibinfo {author} {\bibfnamefont {Nir}\ \bibnamefont
  {Davidson}}, \ and\ \bibinfo {author} {\bibfnamefont {Ofer}\ \bibnamefont
  {Firstenberg}},\ }\bibfield  {title} {\enquote {\bibinfo {title} {Multiport
  atom interferometry for inertial sensing},}\ }\href {\doibase
  10.1103/physreva.100.023617} {\bibfield  {journal} {\bibinfo  {journal}
  {Physical Review A}\ }\textbf {\bibinfo {volume} {100}} (\bibinfo {year}
  {2019}),\ 10.1103/physreva.100.023617}\BibitemShut {NoStop}%
\bibitem [{\citenamefont {Avinadav}\ \emph {et~al.}()\citenamefont {Avinadav},
  \citenamefont {Yankelev}, \citenamefont {Firstenberg},\ and\ \citenamefont
  {Davidson}}]{Avinadav2019}%
  \BibitemOpen
  \bibfield  {author} {\bibinfo {author} {\bibfnamefont {Chen}\ \bibnamefont
  {Avinadav}}, \bibinfo {author} {\bibfnamefont {Dimitry}\ \bibnamefont
  {Yankelev}}, \bibinfo {author} {\bibfnamefont {Ofer}\ \bibnamefont
  {Firstenberg}}, \ and\ \bibinfo {author} {\bibfnamefont {Nir}\ \bibnamefont
  {Davidson}},\ }\bibfield  {title} {\enquote {\bibinfo {title}
  {Composite-fringe atom interferometry for high dynamic-range sensing},}\
  }\href@noop {} {\bibfield  {journal} {\bibinfo  {journal} {arXiv}\ }}\Eprint
  {http://arxiv.org/abs/http://arxiv.org/abs/1912.12304v1}
  {http://arxiv.org/abs/1912.12304v1} \BibitemShut {NoStop}%
\bibitem [{\citenamefont {Lef{\`{e}}vre}(2014)}]{Lefevre2014a}%
  \BibitemOpen
  \bibfield  {author} {\bibinfo {author} {\bibfnamefont {Herv{\'{e}}~C.}\
  \bibnamefont {Lef{\`{e}}vre}},\ }\bibfield  {title} {\enquote {\bibinfo
  {title} {The fiber-optic gyroscope, a century after sagnac's experiment: The
  ultimate rotation-sensing technology?}}\ }\href {\doibase
  10.1016/j.crhy.2014.10.007} {\bibfield  {journal} {\bibinfo  {journal}
  {Comptes Rendus Physique}\ }\textbf {\bibinfo {volume} {15}},\ \bibinfo
  {pages} {851--858} (\bibinfo {year} {2014})}\BibitemShut {NoStop}%
\bibitem [{\citenamefont {Schkolnik}\ \emph {et~al.}(2015)\citenamefont
  {Schkolnik}, \citenamefont {Leykauf}, \citenamefont {Hauth}, \citenamefont
  {Freier},\ and\ \citenamefont {Peters}}]{Schkolnik2015}%
  \BibitemOpen
  \bibfield  {author} {\bibinfo {author} {\bibfnamefont {V.}~\bibnamefont
  {Schkolnik}}, \bibinfo {author} {\bibfnamefont {B.}~\bibnamefont {Leykauf}},
  \bibinfo {author} {\bibfnamefont {M.}~\bibnamefont {Hauth}}, \bibinfo
  {author} {\bibfnamefont {C.}~\bibnamefont {Freier}}, \ and\ \bibinfo {author}
  {\bibfnamefont {A.}~\bibnamefont {Peters}},\ }\bibfield  {title} {\enquote
  {\bibinfo {title} {The effect of wavefront aberrations in atom
  interferometry},}\ }\href {\doibase 10.1007/s00340-015-6138-5} {\bibfield
  {journal} {\bibinfo  {journal} {Applied Physics B}\ }\textbf {\bibinfo
  {volume} {120}} (\bibinfo {year} {2015}),\
  10.1007/s00340-015-6138-5}\BibitemShut {NoStop}%
\bibitem [{\citenamefont {Karcher}\ \emph {et~al.}(2018)\citenamefont
  {Karcher}, \citenamefont {Imanaliev}, \citenamefont {Merlet},\ and\
  \citenamefont {Santos}}]{Karcher2018}%
  \BibitemOpen
  \bibfield  {author} {\bibinfo {author} {\bibfnamefont {R}~\bibnamefont
  {Karcher}}, \bibinfo {author} {\bibfnamefont {A}~\bibnamefont {Imanaliev}},
  \bibinfo {author} {\bibfnamefont {S}~\bibnamefont {Merlet}}, \ and\ \bibinfo
  {author} {\bibfnamefont {F~Pereira~Dos}\ \bibnamefont {Santos}},\ }\bibfield
  {title} {\enquote {\bibinfo {title} {Improving the accuracy of atom
  interferometers with ultracold sources},}\ }\href {\doibase
  10.1088/1367-2630/aaf07d} {\bibfield  {journal} {\bibinfo  {journal} {New
  Journal of Physics}\ }\textbf {\bibinfo {volume} {20}} (\bibinfo {year}
  {2018}),\ 10.1088/1367-2630/aaf07d}\BibitemShut {NoStop}%
\bibitem [{\citenamefont {Kalman}(1960)}]{Kalman1960}%
  \BibitemOpen
  \bibfield  {author} {\bibinfo {author} {\bibfnamefont {R.~E.}\ \bibnamefont
  {Kalman}},\ }\bibfield  {title} {\enquote {\bibinfo {title} {A new approach
  to linear filtering and prediction problems},}\ }\href {\doibase
  10.1115/1.3662552} {\bibfield  {journal} {\bibinfo  {journal} {Journal of
  Basic Engineering}\ }\textbf {\bibinfo {volume} {82}} (\bibinfo {year}
  {1960}),\ 10.1115/1.3662552}\BibitemShut {NoStop}%
\bibitem [{\citenamefont {Cheiney}\ \emph {et~al.}(2018)\citenamefont
  {Cheiney}, \citenamefont {Fouch{\'{e}}}, \citenamefont {Templier},
  \citenamefont {Napolitano}, \citenamefont {Battelier}, \citenamefont
  {Bouyer},\ and\ \citenamefont {Barrett}}]{Cheiney2018}%
  \BibitemOpen
  \bibfield  {author} {\bibinfo {author} {\bibfnamefont {Pierrick}\
  \bibnamefont {Cheiney}}, \bibinfo {author} {\bibfnamefont {Lauriane}\
  \bibnamefont {Fouch{\'{e}}}}, \bibinfo {author} {\bibfnamefont {Simon}\
  \bibnamefont {Templier}}, \bibinfo {author} {\bibfnamefont {Fabien}\
  \bibnamefont {Napolitano}}, \bibinfo {author} {\bibfnamefont {Baptiste}\
  \bibnamefont {Battelier}}, \bibinfo {author} {\bibfnamefont {Philippe}\
  \bibnamefont {Bouyer}}, \ and\ \bibinfo {author} {\bibfnamefont {Brynle}\
  \bibnamefont {Barrett}},\ }\bibfield  {title} {\enquote {\bibinfo {title}
  {Navigation-compatible hybrid quantum accelerometer using a kalman filter},}\
  }\href {\doibase 10.1103/physrevapplied.10.034030} {\bibfield  {journal}
  {\bibinfo  {journal} {Physical Review Applied}\ }\textbf {\bibinfo {volume}
  {10}} (\bibinfo {year} {2018}),\
  10.1103/physrevapplied.10.034030}\BibitemShut {NoStop}%
\bibitem [{\citenamefont {Moral}(1997)}]{Moral1997}%
  \BibitemOpen
  \bibfield  {author} {\bibinfo {author} {\bibfnamefont {Pierre~Del}\
  \bibnamefont {Moral}},\ }\bibfield  {title} {\enquote {\bibinfo {title}
  {Nonlinear filtering: Interacting particle resolution},}\ }\href {\doibase
  10.1016/s0764-4442(97)84778-7} {\bibfield  {journal} {\bibinfo  {journal}
  {Comptes Rendus de l'Acad{\'{e}}mie des Sciences - Series I - Mathematics}\
  }\textbf {\bibinfo {volume} {325}},\ \bibinfo {pages} {653--658} (\bibinfo
  {year} {1997})}\BibitemShut {NoStop}%
\end{thebibliography}%

\end{document}